\documentclass[times]{elsarticle}

\usepackage[T1]{fontenc}
\usepackage{amsmath}
\usepackage{amsfonts}
\usepackage{amssymb}
\usepackage{graphicx}
\usepackage[dvipsnames]{xcolor}
\usepackage{mmacells}
\usepackage{caption}
\usepackage{subcaption}
\usepackage{enumitem}
\usepackage{dsfont}
\usepackage{float}
\usepackage{placeins}
\usepackage{listings}
\usepackage{xcolor}

\newcommand{\BR}{\discretionary{-}{}{}}

\mmaSet{moredefined={LinApart, GaussianIntegers, PreCollect, ApplyAfterPreCollect, GatherByDenominator, Parallel, MultivariateApart}}

\newcommand{\QQ}{\mathbb{Q}}
\newcommand{\CC}{\mathbb{C}}

\biboptions{sort&compress}

\lstdefinelanguage{FORM}{
  alsoletter={.\#},
  morekeywords={
    Symbol,Symbols,Vector,Vectors,Index,Indices,Tensor,Tensors,
    Function,Functions,CFunction,CFunctions,Set,Dimension,AutoDeclare,
    Local,Global,Format,ModuleOption},
  morekeywords=[2]{
    id,identify,repeat,Repeat,endrepeat,EndRepeat,also,multiply,Multiply,
    Print,Bracket,Keep,Collect,Drop,Skip,sum,Trace4,Tracen,Contract,
    Argument,EndArgument,Term,EndTerm,if,else,elseif,endif,while,Discard,
    FillExpression,Hide,Unhide,Inside,EndInside,Transform,PushHide,PopHide,
    nHide,NotInParallel,Symmetrize,ChainIn,ChainOut,SplitArg,FactArg,
    ArgToExtraSymbol,CreateSpectator,CopySpectator,RemoveSpectator,
    ToSpectator,DropCoefficient,PutInside,Repeat},
  morekeywords=[3]{
    .sort,.end,.store,.global,.clear},
  morekeywords=[4]{
    \#define,\#Define,\#include,\#call,\#do,\#Do,\#enddo,\#EndDo,
    \#procedure,\#endprocedure,\#if,\#If,\#else,\#Else,\#endif,\#EndIf,
    \#message,\#write,\#redefine,\#Redefine,\#ifdef,\#IfDef,\#inside,\#Inside,
    \#endinside,\#EndInside},
  morecomment=[f]{*},
  sensitive=true,
  morestring=[b]",
}

\begin{document}
\begin{frontmatter}

\title{{\tt LinApart3}: efficient algorithm for multivariate partial fraction decomposition with linear denominators}

\author[1,2]{L.~Fekésházy\corref{cor1}}
\ead{levente.fekeshazy@desy.de}

\author[3,2]{A.~Kardos}
\ead{kardos.adam@science.unideb.hu}

\cortext[cor1]{Corresponding author}

\affiliation[1]{organization={II.~Institut für Theoretische Physik,
                              Universität Hamburg},
  addressline={Luruper Chaussee 149},
  postcode={22761},
  city={Hamburg},
  country={Germany}}

\affiliation[2]{organization={Institute for Theoretical Physics,
                              ELTE Eötvös Loránd University},
  addressline={Pázmény Péter sétány 1/A},
  postcode={1117},
  city={Budapest},
  country={Hungary}}

\affiliation[3]{organization={Faculty of Science and Technology, Institute of Physics, University of Debrecen},
  addressline={ PO Box 105},
  postcode={H-4010},
  city={Debrecen},
  country={Hungary}}

\begin{abstract}
We present {\tt LinApart3}, an efficient multivariate partial fraction decomposition algorithm for rational functions with linear denominators. Our decomposition algorithm guarantees that each term contains at most as many distinct denominators from the original set as partial fraction variables, introduces no spurious singularities, is independent of variable ordering, and is insensitive to the presence of spectator variables. While general multivariate approaches based on Gröbner-bases or Leinartas' method handle arbitrary polynomial denominators, they suffer from intermediate expression swell. {\tt LinApart3} replaces polynomial-ideal computations with linear algebra and residue extraction by exploiting the geometry of the hyperplane arrangement defined by the denominators, circumventing this issue just as {\tt LinApart} did in the univariate case. Because the individual basis contributions are independent, the algorithm is moreover naturally parallelizable. To showcase the utility of our algorithm we implemented the algorithm both in {\sc Wolfram Mathematica} and \texttt{FORM}.
\end{abstract}


\end{frontmatter}


\newpage
\section*{Program summary}

\begin{itemize}
\item[]{\em Program title:} {\tt LinApart}

\item[]{\em CPC Library link to program files:} (to be assigned)

\item[]{\em Developer's repository link:}
  \url{https://github.com/fekeshazy/LinApart}

\item[]{\em Licensing provisions:} MIT license

\item[]{\em Programming language:} {\sc Wolfram Mathematica}

\item[]{\em Nature of the problem:}
Multivariate rational functions with many denominators occur ubiquitously in perturbative Quantum Field Theory, e.g.\ before and after IBP reduction, in parametric representations, and in analytic phase-space integrations. Standard multivariate partial fraction methods are based on Gröbner bases and polynomial reduction (Leinartas' method); they are general but can be expensive and often obscure the geometric structure. In many physics applications, however, the denominators are linear in the integration variables, defining a hyperplane arrangement. In this case the problem admits a specialized, substantially faster solution.

\item[]{\em Solution method:}
{\tt LinApart3} performs multivariate partial fraction decomposition for
linear denominators using only linear algebra and residue extraction:
\begin{enumerate}[label=(\roman*)]

\item Null-space elimination: 
  Linear dependencies among denominators are found via null spaces of the extended coefficient matrix. After elimination the remaining denominators in each term are linearly independent.
  
\item Basis identification: 
  The length of the basis is the number of partial fraction variables and the elements of the basis are the denominators present in the original expression. A basis is defined as a set of denominators whose coefficient vectors are linearly independent.
  
\item Transformation to denominator space: 
  For each basis, the algorithm performs a coordinate transformation from the space of the partial fraction variables ($\mathbf{x}$) to the space of the basis denominators ($\mathbf{w}$), which is possible due to the fact that the denominators are linear ($w_i = D_{b_i}(\mathbf{x})$).

\item Numerator decomposition: 
  A fraction can be improper, meaning the degree of the numerator in some partial fraction variable can be higher than the degree of said variable in the denominator. In this case the decomposition can contain polynomial terms in said variable. We decompose the expression into a polynomial part and a proper fraction part by going into denominator space in each basis consecutively and expanding the expression.

\item Basis residues in denominator coordinates:
  As the last step, the algorithm computes the partial fraction coefficients via multivariate Taylor expansion at $\mathbf{w}=0$ and adds the contributions together in variable space. Since the contributions from different bases are independent, this step can be parallelized.
  
\end{enumerate}

\item[]{\em Restrictions:}
The multivariate algorithm requires that all variable-dependent denominators are linear in the specified variables.

\end{itemize}

\newpage


\section{Introduction}
\label{sec:intro}

Analytic computations in perturbative Quantum Field Theory frequently produce large rational functions of kinematic invariants and the dimensional regularization parameter. Such expressions arise, for instance, not only in amplitudes before Integration-By-Parts (IBP) reduction and in coefficients of master integrals after IBP  reduction~\cite{Chetyrkin:1981qh,Laporta:2000dsw,Smirnov:2008iw,
Klappert:2020nbg}, but also in parametric representations of Feynman integrals and in the preparation of integrands for analytic integration in terms of multiple polylogarithms~\cite{Goncharov:1998kja,Remiddi:1999ew, Panzer:2014caa,Kardos:2025klp}. Partial fraction decomposition is one of the most effective tools for simplifying these expressions, enabling the detection of cancellations and reducing the complexity of subsequent integration steps.

In the univariate case, partial fraction decomposition is a classical and well-understood problem. Our previous work introduced {\tt LinApart}~\cite{LinApart} for single-variable denominators factorized into linear factors and {\tt LinApart2}~\cite{LinApart2} for single-variable irreducible polynomial denominators of arbitrary degree. Both exploit the Laurent series structure of univariate rational functions to bypass the equation-system approach used by standard tools such as Mathematica's built-in {\tt Apart}, achieving substantial runtime-speedups and reduced memory usage.

The multivariate case, however, presents fundamental new challenges. A naive iterated application of univariate partial fraction decomposition introduces spurious denominators: factors that are not present in the original expression and arise from the decomposition process. These spurious poles complicate subsequent analysis, can obscure cancellations, and produce apparent singularities that vanish only after recombination. Moreover, the result of iterated univariate decomposition
depends on the order in which the variables are processed.

A mathematically more rigorous framework for multivariate partial fraction decomposition was provided by Leinartas~\cite{Leinartas}, who showed that any rational function can be written as a sum of terms like
\begin{equation}
    \label{eq:leinartas-form}
    f(\mathbf{x}) = 
                    \sum_A \frac{N_A(\mathbf{x})}{\prod_{i\in A}
                    D_i(\mathbf{x})^{a_i}}\,,
\end{equation}
where the sum runs over subsets $A$ of the original set of denominators, and each subset must satisfy two conditions. First, the denominators in a given term must share a common zero; if no such point exists, the term can be split further using the identity $1 = \sum_{i \in A} g_i\, D_i$. Second, the denominators in each term must be algebraically independent, which in~$n$ variables limits each term to at most~$n$ denominator factors. Geometrically, this means the zero sets of the denominators intersect at an isolated point rather than along a higher-dimensional subspace. By construction, this decomposition involves only denominators already present in the original expression. Algorithmic realizations of Leinartas' decomposition have been published for example by Raichev~\cite{Raichev:2012}. 

Further progress was made  by Heller and von Manteuffel in {\tt Multivariate\BR Apart}~\cite{MultivariateApart}.  They reformulated the decomposition as a polynomial reduction problem with respect to an ideal, solved via Gr\"obner basis computation. While these algorithms are general and mathematically rigorous they are resource intensive. Take for example the Gröbner-basis step, which has worst-case doubly-exponential complexity and can produce significant intermediate expression swell, particularly when the coefficient ring contains many symbolic parameters, as is typical in physics applications.

In many cases of practical interest, however, the denominators possess additional structure that can be exploited. Propagators in Feynman parameter space, linear cuts in phase-space integrals, Mellin-Barnes integration variables, and denominators arising from IBP reduction are almost exclusively linear in the relevant variables. For such denominators, the set $\{D_i(\mathbf{x})\}$ defines a hyperplane arrangement in~$\CC^n$, and Leinartas' conditions reduce to linear algebra:
\begin{enumerate}
    \item condition~(i), the existence of a common zero, becomes consistency of the affine system $\{D_i=0\}$, which fails when an affine relation $\sum_i \alpha_i D_i = \mathrm{const} \neq 0$ holds (parallel hyperplanes), testable by comparing the ranks of the coefficient matrix and its extension by the constant column,
    
    \item condition~(ii) becomes algebraic independence of the denominators, which for linear forms is linear independence of their extended coefficient vectors $(c_{i1},\ldots,c_{in},d_i)$; its failure is a null relation, a linear combination of denominators that vanishes identically, detectable via the null space of the coefficient matrix.
\end{enumerate}
In this paper we present {\tt LinApart3}, which exploits this reduction to linear algebra. The algorithm proceeds in two stages. First, null relations among the denominators are identified and used to construct one-operators that recursively eliminate linearly dependent denominators. Second, for each linearly independent $n$-element subset of denominators (a ``basis''), a coordinate transformation to denominator space is performed and the partial fraction coefficients are extracted via a multivariate residue formula; the direct generalization of the Laurent series residues used in {\tt LinApart} and {\tt LinApart2}. The algorithm also handles general rational functions with polynomial numerators of arbitrary degree: improper fractions are separated into a polynomial part and a proper fractional part, and variable-dependent numerators are expanded in denominator space before the residue computation. The basis contributions are independent of one another and can be computed in parallel.

The remainder of this paper is organized as follows. Section~\ref{sect:algorithm} describes the algorithm in detail with its mathematical foundations. Section~\ref{sect:example} presents a complete worked example. Section~\ref{sect:Mathematica} discusses the {\sc Wolfram Mathematica} implementation. Section~\ref{sect:benchmarks} provides performance benchmarks among the algorithms implemented in the {\sc Wolfram Mathematica} language.  Section~\ref{sect:FORM} discusses the \texttt{FORM}
implementation and a benchmark in a standard real life example. Section~\ref{sect:conclusions} contains our conclusions.

\section{Algorithm}
\label{sect:algorithm}

Consider a rational function of $n$ variables $\mathbf{x} = (x_1, \ldots, x_n)$ with $k$ linear denominators,
\begin{equation}
    \label{eq:input}
    f(\mathbf{x}) = \frac{P(\mathbf{x})}
    {\prod_{i=1}^{k} D_i(\mathbf{x})^{m_i}}\,,
\end{equation}
where $P(\mathbf{x})$ is a polynomial and each denominator is a linear form
\begin{equation}
    \label{eq:linear_form}
    D_i(\mathbf{x}) = \sum_{j=1}^{n} c_{ij}\, x_j + d_i\,,
\end{equation}
with coefficients $c_{ij}$ and $d_i$ that may depend on additional parameters but are independent of $\mathbf{x}$. We encode these denominators in the extended coefficient matrix
\begin{equation}
    \label{eq:extended_matrix}
    \mathcal{C} = \begin{pmatrix}
    c_{11} & \cdots & c_{1n} & d_1 \\
    \vdots & \ddots & \vdots & \vdots \\
    c_{k1} & \cdots & c_{kn} & d_k
    \end{pmatrix} \in \QQ^{k \times (n+1)}\,.
\end{equation}
The algorithm proceeds in four steps: elimination of null relations, identification of bases, expansion of variable-dependent numerators, and extraction of basis residues. We describe each in turn.

\subsection{Null relation elimination}
\label{subsect:null_elimination}

A null-relation is a nontrivial vector $\boldsymbol{\alpha} = (\alpha_1, \ldots, \alpha_k)$ satisfying
\begin{equation}
    \label{eq:null_relation}
    \sum_{i=1}^{k} \alpha_i\, D_i(\mathbf{x}) = 0
    ~\text{for all}~\mathbf{x}\,.
\end{equation}
Such a relation exists if and only if the rows of $\mathcal{C}$ indexed by the support of $\boldsymbol{\alpha}$ are linearly dependent. We only need the minimal null relations, those whose support $S$ becomes linearly independent upon removing any single element. Since the extended rows live in $\QQ^{n+1}$, deleting one element of a minimal relation leaves at most $n+1$ independent vectors, so $|S|-1 \le n+1$ and a minimal null relation involves at most $n+2$ forms. It therefore suffices to compute the null space of $\mathcal{C}_S^T$ for all subsets $S \subseteq \{1,\ldots,k\}$ with $3 \le |S| \le n+2$; all two elements subsets represent equalities between denominators, which by construction cannot happen. Restricting to minimal relations loses nothing: every linearly dependent set of denominators contains such a minimal relation, so each term that still carries a dependency is reduced by one of them.

Each null relation provides a way to reduce the number of distinct denominators in a term by one. Given a relation $\sum_{i \in S} \alpha_i\, D_i = 0$, we select one denominator $D_\ell$ with $\ell \in S$ and solve for it:
\begin{equation}
    \label{eq:solve_for_Dl}
    D_\ell = -\frac{1}{\alpha_\ell}
    \sum_{\substack{i \in S \\ i \neq \ell}} \alpha_i\, D_i\,.
\end{equation}
This allows us to construct the one-operator
\begin{equation}
    \label{eq:one_operator}
    1= \frac{1}{D_\ell} \left(
    -\frac{1}{\alpha_\ell}
    \sum_{\substack{i \in S \\ i \neq \ell}} \alpha_i\, D_i
    \right)\,,
\end{equation}
which is identically equal to unity by eq.~\eqref{eq:solve_for_Dl}. Multiplying a term by $1$ and expanding enables cancellation of $D_i$s, with the tradeoff of increasing the multiplicity of $D_\ell$.

The elimination proceeds recursively: at each step null-relations are identified and one is chosen. The corresponding one-operator is applied, and the resulting terms are processed independently. The recursion terminates when no further null relations exist among the denominators of any term, meaning that in each term all remaining denominators are linearly independent. 

The choice of $D_\ell$ and the null-relation at each step is arbitrary. However, an adaptive (per-step) choice of $D_{\ell}$ can lead to infinite loops. Furthermore, the choice of the null-relation can influence the number of remaining additive terms. In order to get a terminating algorithm and avoid oscillation, we choose the denominator to be eliminated according to a priority ordering by multiplicity (highest first in the {\sc Mathematica} implementation; the \texttt{FORM} implementation uses the reverse) fixed before the recursion begins. We use this ordering to avoid many high multiplicities and to minimize the number of terms. This avoids a large number of higher-order residue computations and a large number of bases, both of which strongly influence the runtime.

\subsection{Basis identification}
\label{subsect:basis_identification}

After null relation elimination, each term contains only linearly independent denominators. However, a term may still contain more than~$n$ distinct denominator factors, and these must be decomposed further into contributions from $n$-element bases.
 
We call such $n$-subset of the denominators present in a given term a basis ($\mathcal{B} = \{D_{b_1}, \ldots, D_{b_n}\}$) whose $n \times n$ coefficient matrix
\begin{equation}
    \label{eq:basis_coeff_matrix}
    C_{\mathcal{B}} =
        \begin{pmatrix}
        c_{b_1,1} & \cdots & c_{b_1,n} \\
        \vdots & \ddots & \vdots \\
        c_{b_n,1} & \cdots & c_{b_n,n}
        \end{pmatrix}
    \neq
    0
\end{equation}
has nonzero determinant, where $c_{b_i,j}$ are the coefficients of~$x_j$ in~$D_{b_i}$. Geometrically, this means that the $n$ hyperplanes $\{D_{b_i}(\mathbf{x}) = 0\}$ intersect at a single point in~$\CC^n$, rather than along a higher-dimensional subspace or being parallel.

\subsection{Numerator expansion in denominator space}
\label{subsect:numerator_expansion}

After null relation elimination, some terms may have numerators that depend on the variables $\mathbf{x}$. These must be reduced to constant numerators before the residue extraction of the next
step. We choose to do this here even though it could theoretically also be done before null-relation elimination for two reasons. First, it requires an independent set of denominators, which can be troublesome to determine before eliminating the linear dependencies between the denominators. Second, it would introduce a new arbitrary choice, which would influence the decomposition.

We introduce denominator coordinates $\mathbf{w} = (w_1, \ldots, w_n)$ defined by $w_i = B_i(\mathbf{x})$. Since $\det C_B \neq 0$, this system can be inverted to give $\mathbf{x} = \mathbf{x}(\mathbf{w})$. In these coordinates any polynomial numerator $P(\mathbf{x})$ becomes a polynomial in $\mathbf{w}$:
\begin{equation}
    P(\mathbf{x}(\mathbf{w})) =
    \sum_{\boldsymbol{\beta}} p_{\boldsymbol{\beta}}\,
    w_1^{\beta_1} \cdots w_n^{\beta_n}
    \,,\quad
    \boldsymbol{\beta} = (\beta_1,\dots,\beta_n)\,.
\end{equation}
Each monomial $w_i^{\beta_i}$ cancels against the pole $1/w_i^{m_i} = 1/B_i^{m_i}$, effectively reducing the multiplicity. After transforming back to the original variables, the numerator becomes independent of $\mathbf{x}$.

This expansion is performed for every basis before the residue computation to ensure that each term entering the next step has a constant numerator.

\subsection{Basis residues}
\label{subsect:basis_residues}

After the first two steps, each term has linearly independent denominators and a constant numerator. The final step decomposes each such term into a sum of contributions, one for each basis.

A set of $n$ linearly independent denominators $\{B_1, \ldots, B_n\}$ defines a unique point
$\mathbf{x}^* \in \CC^n$ at which all $n$ hyperplanes intersect: $B_1(\mathbf{x}^*) = \cdots = B_n(\mathbf{x}^*) = 0$. The partial fraction contribution from this intersection is obtained
by computing the multivariate residue at $\mathbf{x}^*$.

The coordinate transformation $w_i = B_i(\mathbf{x})$ plays a central role: it maps the intersection point to the origin $\mathbf{w} = 0$ and simultaneously diagonalizes the basis denominators, so that $B_i = w_i$. In these coordinates, a term with basis multiplicities $(m_1, \ldots, m_n)$ and a generating function $\mathcal{G}(\mathbf{w})$ encoding the non-basis denominators and the coefficient takes the form
\begin{equation}
    \label{eq:term_in_w}
    \frac{\mathcal{G}(\mathbf{w})}{w_1^{m_1} \cdots w_n^{m_n}}\,.
\end{equation}
The partial fraction coefficients are then given by the Grothendieck residue at the origin. For a meromorphic $n$-form $\omega = g(\mathbf{w})\, dw_1 \wedge \cdots \wedge dw_n\,/\, (w_1^{m_1} \cdots w_n^{m_n})$, this residue is defined as the iterated contour integral
\begin{equation}
    \label{eq:grothendieck}
    \mathrm{Res}_{\mathbf{w}=0}\,\omega =
    \frac{1}{(2\pi i)^n}
    \oint_{|w_1|=\epsilon_1} \cdots \oint_{|w_n|=\epsilon_n}
    \frac{g(\mathbf{w})}{w_1^{m_1} \cdots w_n^{m_n}}\,
    dw_1 \cdots dw_n\,.
\end{equation}
Because the denominator is a product of powers of the coordinate variables, the iterated integral factorizes: each contour integral in $w_i$ independently extracts a Laurent coefficient via the Cauchy formula. The result is
\begin{equation}
    \label{eq:residue_formula}
    \mathrm{Res}_{\mathbf{w}=0}\,\omega =
    \frac{1}{\prod_{i=1}^n (m_i-1)!}
    \left[\frac{\partial^{m_1-1}}{\partial w_1^{m_1-1}} \cdots
    \frac{\partial^{m_n-1}}{\partial w_n^{m_n-1}}\,
    g(\mathbf{w})\right]_{\mathbf{w}=0}\,.
\end{equation}
This is the direct multivariate generalization of the Laurent series residue formula used in {\tt LinApart} and {\tt LinApart2}: there, the coordinate shift $w = x - a$ moves a univariate pole to the origin and residues are extracted by differentiation; here, the transformation $w_i = B_i(\mathbf{x})$ does the same in $n$ dimensions simultaneously.

To obtain the full partial fraction decomposition, we need not only the leading residue but all Laurent coefficients. Expanding eq.~\eqref{eq:term_in_w} gives
\begin{equation}
    \label{eq:laurent_expansion}
    \frac{\mathcal{G}(\mathbf{w})}{w_1^{m_1} \cdots w_n^{m_n}} =
    \sum_{j_1=1}^{m_1} \cdots \sum_{j_n=1}^{m_n}
    \frac{c_{j_1,\ldots,j_n}}{w_1^{j_1} \cdots w_n^{j_n}}
    + \text{regular terms}\,,
\end{equation}
where each coefficient is itself a Grothendieck residue,
\begin{equation}
\label{eq:coefficient_formula}
    c_{j_1,\ldots,j_n} =
    \frac{1}{\prod_{i=1}^n (m_i - j_i)!}
    \left[\frac{\partial^{m_1-j_1}}{\partial w_1^{m_1-j_1}} \cdots
    \frac{\partial^{m_n-j_n}}{\partial w_n^{m_n-j_n}}\,
    \mathcal{G}(\mathbf{w})\right]_{\mathbf{w}=0}\,.
\end{equation}
Transforming back to the original variables via $w_i = B_i$, the contribution from a single basis is
\begin{equation}
    \label{eq:basis_contribution}
    \sum_{j_1=1}^{m_1} \cdots \sum_{j_n=1}^{m_n}
    \frac{c_{j_1,\ldots,j_n}}{B_1^{j_1} \cdots B_n^{j_n}}\,.
\end{equation}
The complete decomposition is the sum of eq.~\eqref{eq:basis_contribution} over all bases and can be regarded as a multivariate Laurent-series. We would like to emphasize that since each basis contribution depends only on its own generating function, these computations are independent and can be performed in parallel.


\section{Worked example}
\label{sect:example}

We would like to illustrate the algorithm on a very simple rational function:
\begin{equation}
    \label{eq:example_input}
    f(x,y) = \frac{1}{x^2\, y\, (x+y)\, (x+y-1)}\,,
\end{equation}
which has $n=2$ variables and $k=4$ linear denominators
\begin{equation}
    D_1 = x\,,\quad D_2 = y\,,\quad D_3 = x+y\,,\quad D_4 = x+y-1\,,
\end{equation}
with multiplicities $m_1=2$, $m_2=m_3=m_4=1$.

We acknowledge that this example could be decomposed by hand more easily with other methods, but our goal here is to explain the algorithm rather than to demonstrate its efficiency.
 
\subsection{Step 1: Null relation elimination}
 
The extended coefficient matrix is
\begin{equation}
    \mathcal{C} =
        \begin{pmatrix}
        1 & 0 & 0 \\
        0 & 1 & 0 \\
        1 & 1 & 0 \\
        1 & 1 & -1
        \end{pmatrix}\,.
\end{equation}
Searching over subsets of size $3 \le |S| \le n+2 = 4$, we find a single null relation
\begin{equation}
    \label{eq:example_null}
    D_1 + D_2 - D_3 = 0\,,
\end{equation}
corresponding to $\boldsymbol{\alpha} = (1,1,-1,0)$. No subset of size~$2$ yields a null relation (no two denominators are proportional), and the relation $D_3 - D_4 = 1 \neq 0$ is an affine but not a null relation: the constant term does not vanish, so it cannot be used for elimination.
 
The priority ordering is determined by multiplicity, highest first. The multiplicities are $(m_1, m_2, m_3, m_4) = (2,1,1,1)$, so the priority list is $(D_1, D_2, D_3, D_4)$. Among the denominators participating in eq.~\eqref{eq:example_null}, the highest-priority one is~$D_1$. We therefore choose to eliminate~$D_1$.
 
Solving for $D_1$ gives $D_1 = D_3 - D_2$, and the one-operator is
\begin{equation}
    1 = \frac{D_3 - D_2}{D_1}\,.
\end{equation}
Multiplying $f$ by 1 and expanding, we obtain
\begin{align}
f &= \frac{1}{D_1^2\, D_2\, D_3\, D_4}
     \cdot \frac{D_3 - D_2}{D_1} \nonumber\\
  &= \frac{D_3}{D_1^3\, D_2\, D_3\, D_4}
   - \frac{D_2}{D_1^3\, D_2\, D_3\, D_4} \nonumber\\
  &= \frac{1}{x^3\, y\, (x{+}y{-}1)}
   - \frac{1}{x^3\, (x{+}y)\, (x{+}y{-}1)}\,.
\label{eq:first_elimination}
\end{align}
The first term contains three distinct denominators $\{D_1, D_2, D_4\}$ and the second contains $\{D_1, D_3, D_4\}$. In both cases the extended coefficient matrix has full rank, so no further null relation exists. The price we paid is an increase in the multiplicity of~$D_1$ from~$2$ to~$3$. Note that the total denominator multiplicity is conserved term by term: the one-operator raises the multiplicity of $D_{\ell}$ by exactly the number of factors it removes. The elimination is complete after a single step, and we have
\begin{equation}
    \label{eq:after_elimination}
    f = 
        \underbrace{\frac{1}{x^3\, y\, (x{+}y{-}1)}}_{T_A}
        \underbrace{- \frac{1}{x^3\, (x{+}y)\, (x{+}y{-}1)}}_{T_B}\,.
\end{equation}
Each term has linearly independent denominators and a constant numerator, so step~2 (numerator expansion) is trivial and we proceed directly to step~3.
 
\subsection{Step 2: Basis identification}
 
We enumerate the valid two-element bases for each term by checking which pairs of denominators have linearly independent coefficient vectors, i.e.\ nonzero determinant of the $2\times 2$ coefficient matrix (excluding the constant column).
 
For term~$T_A$ the denominators are $\{D_1, D_2, D_4\}$:
\begin{gather}
\{D_1, D_2\}:\;
\det\!\begin{pmatrix} 1 & 0 \\ 0 & 1 \end{pmatrix} = 1 \neq 0\,,
\qquad
\{D_1, D_4\}:\;
\det\!\begin{pmatrix} 1 & 0 \\ 1 & 1 \end{pmatrix} = 1 \neq 0\,, \\ \nonumber
\qquad
\{D_2, D_4\}:\;
\det\!\begin{pmatrix} 0 & 1 \\ 1 & 1 \end{pmatrix} = -1 \neq 0\,.
\end{gather}
All three pairs form valid bases.

For term~$T_B$ the denominators are $\{D_1, D_3, D_4\}$:
\begin{gather}
\{D_1, D_3\}:\;
\det\!\begin{pmatrix} 1 & 0 \\ 1 & 1 \end{pmatrix} = 1 \neq 0\,,
\qquad
\{D_1, D_4\}:\;
\det\!\begin{pmatrix} 1 & 0 \\ 1 & 1 \end{pmatrix} = 1 \neq 0\,, \\ \nonumber
\qquad
\{D_3, D_4\}:\;
\det\!\begin{pmatrix} 1 & 1 \\ 1 & 1 \end{pmatrix} = 0\,.
\end{gather}
The pair $\{D_3, D_4\}$ is not a basis because the two hyperplanes
$x+y = 0$ and $x+y-1 = 0$ are parallel.
 
\subsection{Step 3: Basis residues}
 
\subsubsection{Term $T_A$: $\dfrac{1}{x^3\,y\,(x{+}y{-}1)}$}
\hspace*{0.5cm}
 
The denominators are $D_1 = x$, $D_2 = y$, $D_4 = x+y-1$ with multiplicities $(3,1,1)$.
 
\paragraph{Basis $\{D_1, D_2\}$ with multiplicities $(3,1)$}

The denominator coordinates are $w_1 = x$, $w_2 = y$, which gives $x = w_1$, $y = w_2$. The non-basis denominator transforms as $D_4 = x+y-1 = w_1+w_2-1$, giving the generating function
\begin{equation}
    \mathcal{G}(\mathbf{w}) = \frac{1}{w_1+w_2-1}\,.
\end{equation}
Since $m_2 = 1$, only $j_2 = 1$ contributes. The coefficients for $j_1 = 1,2,3$ are computed via eq.~\eqref{eq:coefficient_formula}:
\begin{align}
    c_{3,1} &= \frac{1}{0!\,0!}
      \left[\frac{1}{w_1{+}w_2{-}1}\right]_{\mathbf{w}=0}
      = -1\,, \\
    c_{2,1} &= \frac{1}{1!\,0!}
      \left[\frac{\partial}{\partial w_1}\frac{1}{w_1{+}w_2{-}1}
      \right]_{\mathbf{w}=0}
      = \frac{-1}{(-1)^2} = -1\,, \\
    c_{1,1} &= \frac{1}{2!\,0!}
      \left[\frac{\partial^2}{\partial w_1^2}
      \frac{1}{w_1{+}w_2{-}1}\right]_{\mathbf{w}=0}
      = \frac{1}{2}\cdot\frac{2}{(-1)^3} = -1\,.
\end{align}
The contribution from this basis is
\begin{equation}
    \label{eq:TA_basis12}
    -\frac{1}{x^3 y} - \frac{1}{x^2 y} - \frac{1}{x y}\,.
\end{equation}
 
\paragraph{Basis $\{D_1, D_4\}$ with multiplicities $(3,1)$}

The denominator coordinates are $w_1 = x$, $w_2 = x+y-1$, giving $x = w_1$, $y = w_2 - w_1 + 1$. The non-basis denominator is $D_2 = y = w_2 - w_1 + 1$ with multiplicity~1, so
\begin{equation}
    \mathcal{G}(\mathbf{w}) = \frac{1}{w_2 - w_1 + 1}\,.
\end{equation}
Again only $j_2 = 1$ contributes. The coefficients for $j_1 = 1,2,3$ are:
\begin{align}
    c_{3,1} &= \frac{1}{0!\,0!}
      \left[\frac{1}{w_2{-}w_1{+}1}\right]_{\mathbf{w}=0}
      = 1\,, \\
    c_{2,1} &= \frac{1}{1!\,0!}
      \left[\frac{\partial}{\partial w_1}
      \frac{1}{w_2{-}w_1{+}1}\right]_{\mathbf{w}=0}
      = \frac{1}{(1)^2} = 1\,, \\
    c_{1,1} &= \frac{1}{2!\,0!}
      \left[\frac{\partial^2}{\partial w_1^2}
      \frac{1}{w_2{-}w_1{+}1}\right]_{\mathbf{w}=0}
      = \frac{1}{2}\cdot\frac{2}{(1)^3} = 1\,,
\end{align}
The contribution is
\begin{equation}
    \label{eq:TA_basis14}
    \frac{1}{x^3(x{+}y{-}1)} + \frac{1}{x^2(x{+}y{-}1)}
    + \frac{1}{x(x{+}y{-}1)}\,.
\end{equation}
 
\paragraph{Basis $\{D_2, D_4\}$ with multiplicities $(1,1)$}

The denominator coordinates are $w_1 = y$, $w_2 = x+y-1$, giving $y = w_1$, $x = w_2 - w_1 + 1$. The non-basis denominator is $D_1 = x = w_2 - w_1 + 1$ with multiplicity~3, so
\begin{equation}
    \mathcal{G}(\mathbf{w}) = \frac{1}{(w_2 - w_1 + 1)^3}\,.
\end{equation}
The single coefficient is
\begin{equation}
c_{1,1} = \frac{1}{0!\,0!}
  \left[\frac{1}{(w_2{-}w_1{+}1)^3}\right]_{\mathbf{w}=0}
  = 1\,.
\end{equation}
The contribution is
\begin{equation}
    \label{eq:TA_basis24}
    \frac{1}{y(x{+}y{-}1)}\,.
\end{equation}
 
Combining eqs.~\eqref{eq:TA_basis12}, \eqref{eq:TA_basis14}, and~\eqref{eq:TA_basis24},
\begin{equation}
    \label{eq:TA_result}
    T_A = -\frac{1}{x^3 y} - \frac{1}{x^2 y} - \frac{1}{x y}
          + \frac{1}{x^3(x{+}y{-}1)} + \frac{1}{x^2(x{+}y{-}1)}
          + \frac{1}{x(x{+}y{-}1)} + \frac{1}{y(x{+}y{-}1)}\,.
\end{equation}

\subsubsection{Term $T_B$: $-\dfrac{1}{x^3\,(x{+}y)\,(x{+}y{-}1)}$}
\hspace*{0.5cm}
 
The denominators are $D_1 = x$, $D_3 = x+y$, $D_4 = x+y-1$ with multiplicities $(3,1,1)$. We carry the overall minus sign into the generating function.
 
\paragraph{Basis $\{D_1, D_3\}$ with multiplicities $(3,1)$}

Denominator coordinates: $w_1 = x$, $w_2 = x+y$, so $x = w_1$, $y = w_2 - w_1$. The non-basis denominator is $D_4 = x+y-1 = w_2 - 1$:
\begin{equation}
    \mathcal{G}(\mathbf{w}) = \frac{-1}{w_2-1}\,.
\end{equation}
Since $\mathcal{G}$ is independent of $w_1$, all derivatives with respect to~$w_1$ vanish, and only $j_1 = m_1 = 3$ contributes:
\begin{equation}
    c_{3,1} = \frac{1}{0!\,0!}
      \left[\frac{-1}{w_2-1}\right]_{\mathbf{w}=0}
      = 1\,.
\end{equation}
The contribution from this basis is
\begin{equation}
\label{eq:TB_basis13}
\frac{1}{x^3(x{+}y)}\,.
\end{equation}
 
\paragraph{Basis $\{D_1, D_4\}$ with multiplicities $(3,1)$}

Denominator coordinates: $w_1 = x$, $w_2 = x+y-1$, so $x = w_1$, $y = w_2 - w_1 + 1$. The non-basis denominator is $D_3 = x+y = w_2 + 1$:
\begin{equation}
    \mathcal{G}(\mathbf{w}) = \frac{-1}{w_2+1}\,.
\end{equation}
Again $\mathcal{G}$ is independent of~$w_1$, so only $j_1 = 3$ contributes:
\begin{equation}
    c_{3,1} = \frac{1}{0!\,0!}
      \left[\frac{-1}{w_2+1}\right]_{\mathbf{w}=0}
      = -1\,.
\end{equation}
The contribution is
\begin{equation}
    \label{eq:TB_basis14}
    -\frac{1}{x^3(x{+}y{-}1)}\,.
\end{equation}
 
Combining eqs.~\eqref{eq:TB_basis13} and~\eqref{eq:TB_basis14},
\begin{equation}
    \label{eq:TB_result}
    T_B = \frac{1}{x^3(x{+}y)}
          - \frac{1}{x^3(x{+}y{-}1)}\,.
\end{equation}
 
\subsection{Final result}
 
Summing eqs.~\eqref{eq:TA_result} and~\eqref{eq:TB_result}, we observe that the terms $\pm 1/(x^3(x{+}y{-}1))$ cancel between $T_A$ and $T_B$. The complete partial fraction decomposition is
\begin{align}
    \label{eq:example_result}
    \frac{1}{x^2\, y\, (x{+}y)\, (x{+}y{-}1)}
    &= -\frac{1}{x^3 y} - \frac{1}{x^2 y} - \frac{1}{x y}
       + \frac{1}{x^2(x{+}y{-}1)} + \frac{1}{x(x{+}y{-}1)} \nonumber\\
    &\quad + \frac{1}{y(x{+}y{-}1)}
       + \frac{1}{x^3(x{+}y)}\,.
\end{align}
Every term contains at most $n=2$ denominators, all drawn from the original set $\{x,\, y,\, x{+}y,\, x{+}y{-}1\}$; thus no spurious denominators have been introduced.


\section{Usage}
\label{sect:Mathematica}

In this section, we present the usage of {\tt LinApart3} in the Wolfram Mathematica language, which is publicly available at \url{https://github.com/fekeshazy/LinApart}. 

The multivariate functionality is fully integrated into the existing {\tt LinApart} package, preserving the user interface of the previous versions while introducing three new multivariate decomposition methods.

The package can be loaded in any session by specifying the complete path to the file when loading,
\begin{mmaCell}{Input}
  Import["/path/to/LinApart/Mathematica/LinApart.m"]
\end{mmaCell}

\subsection{Multivariate residue method}

The multivariate partial fraction decomposition with respect to the variables {\{\tt x,y,\ldots \}} is invoked by passing a list of variables as the second argument,
\begin{mmaCell}{Input}
  LinApart[1/(x y (x + y - 1)), \{x, y\}]
\end{mmaCell}
\begin{mmaCell}{Output}
  -1/(x y) + 1/(x (-1 + x + y)) + 1/(y (-1 + x + y))
  
\end{mmaCell}

As shown above, if a null relation exists among the denominators, the algorithm eliminates the linearly dependent denominators and then computes the residues. Taking the example of Section~\ref{sect:example} for instance,
\begin{mmaCell}{Input}
  LinApart[1/(x^2 y (x + y) (x + y - 1)), \{x, y\}]
\end{mmaCell}
\begin{mmaCell}{Output}
  -(1/(x^3 y)) - 1/(x^2 y) - 1/(x y) + 1/(x^2 (-1 + x + y)) + 
  1/(x (-1 + x + y)) + 1/(y (-1 + x + y)) + 1/(x^3 (x + y))
  
\end{mmaCell}

The algorithm extends to any number of variables. The output is guaranteed to contain at most~$n$ denominator factors per term,
\begin{mmaCell}{Input}
  LinApart[1/(x y z (x + y + z - 1)), \{x, y, z\}]
\end{mmaCell}
\begin{mmaCell}{Output}
  -(1/(x y z)) + 1/(x y (-1 + x + y + z)) + 
  1/(x z (-1 + x + y + z)) + 1/(y z (-1 + x + y + z))
  
\end{mmaCell}

The denominator coefficients may depend on additional symbolic parameters. Such parameters are treated as elements of the coefficient ring and do not participate in the decomposition,
\begin{mmaCell}{Input}
  LinApart[1/((s - x)(t - y)(s + t - x - y - 1)), \{x, y\}]
\end{mmaCell}
\begin{mmaCell}{Output}
  -(1/((s - x) (t - y))) + 1/((s - x) (-1 + s + t - x - y)) + 
  1/((t - y) (-1 + s + t - x - y))
  
\end{mmaCell}

Expressions with polynomial numerators are handled automatically. The numerator is expanded in denominator space (cf.\ section~\ref{subsect:numerator_expansion}),
\begin{mmaCell}{Input}
  LinApart[(x + 2 y)/((x - y) y (x + y - 1)), \{x, y\}]
\end{mmaCell}
\begin{mmaCell}{Output}
  3/((x - y) (-1 + x + y)) + 1/(y (-1 + x + y))
\end{mmaCell}
Improper fractions, where the numerator degree exceeds the total denominator degree, are also handled. A polynomial part (in said variable) is extracted before the fractional part is decomposed,
\begin{mmaCell}{Input}
  LinApart[x^3/((x - y) y (x + y - 1)), \{x, y\}]//Expand
\end{mmaCell}
\begin{mmaCell}{Output}
  -(1/4) + 1/(4 (x - y)) + x/(4 (x - y)) + 1/y + x/y + 
  y/(4 (x - y)) - 5/(4 (-1 + x + y)) + 
  1/(4 (x - y) (-1 + x + y)) + 
  1/(y (-1 + x + y)) + y/(2 (-1 + x + y))
     
\end{mmaCell}

The multivariate residue method requires all variable-dependent denominators to be linear. Non-linear factors are detected by the preprocessor, factored out, and carried through multiplicatively without participating in the decomposition,
\begin{mmaCell}{Input}
  LinApart[1/((x^2 + y^2)(x + 1)(y - 2)), \{x, y\}]
\end{mmaCell}
\begin{mmaCell}{Message}
  LinApart::nonLinearDenomFactored: Warning: 
  Non-linear denominators \{x^2+y^2\} were factored out and not decomposed.
  
\end{mmaCell}
\begin{mmaCell}{Output}
  1/((1 + x) (-2 + y) (x^2 + y^2))
\end{mmaCell}
A diagnostic message informs the user that non-linear factors were not decomposed. For expressions with non-linear denominators, the Leinartas or Gröbner methods may be used instead.

\subsection{Method selection}

For multivariate partial fraction decomposition three multivariate methods are available via the {\tt "Method"} option. By default, the multivariate residue method is used when the variable argument is a list. However that is restricted to linear denominators as shown previously.

In order to remedy this and be able to benchmark our method with the most used algorithms, we have implemented the Leinartas' decomposition method via syzygy elimination~\cite{Leinartas,
Raichev:2012} and the Gröbner basis method following the approach of ref.~\cite{MultivariateApart}. These are able to handle arbitrary degree multivariate polynomial denominators as well.

\begin{mmaCell}{Input}
  LinApart[1/((x^2 + y^2)(x + y)(x - y + 1)), \{x, y\},
    "Method" -> "Leinartas"]
    
\end{mmaCell}
\begin{mmaCell}{Output}
  2/((1 + x - y) (x + y)) - x/((1 + x - y) (x^2 + y^2)) - 
  y/((1 + x - y) (x^2 + y^2)) + 1/((x + y) (x^2 + y^2)) - 
  x/((x + y) (x^2 + y^2)) + y/((x + y) (x^2 + y^2))
  
\end{mmaCell}
\begin{mmaCell}{Input}
  LinApart[1/((x^2 + y^2)(x + y)(x - y + 1)), \{x, y\},
    "Method" -> "Groebner"]
    
\end{mmaCell}
\begin{mmaCell}{Output}
  2/((1 + x - y) (x + y)) - 2/(x^2 + y^2) + 
  1/((1 + x - y) (x^2 + y^2)) - 
  (2 y)/((1 + x - y) (x^2 + y^2)) + 
  1/((x + y) (x^2 + y^2)) + (2 y)/((x + y) (x^2 + y^2))
  
\end{mmaCell}

Due to the fundamental mathematical differences between the methods, they produce significantly different outputs; however, all are valid partial fraction decompositions. We intentionally do not simplify the  output in all cases and leave it up to the user to structure their expression in the most suitable way they would like.

Consider the following example with linear denominators, where all three methods produce equivalent but structurally different results,
\begin{mmaCell}{Input}
  \mmaDef{expr} = x^2/(y (x + y)(x + y - 1));
  LinApart[\mmaDef{expr}, \{x, y\}]
  LinApart[\mmaDef{expr}, \{x, y\}, "Method" -> "Leinartas"]
  LinApart[\mmaDef{expr}, \{x, y\}, "Method" -> "Groebner"]
  
\end{mmaCell}
\begin{mmaCell}{Output}
  1/y + 1/(y (-1 + x + y)) - 2/(x + y) - 
  2/((-1 + x + y) (x + y)) + y/((-1 + x + y) (x + y))
  
\end{mmaCell}
\begin{mmaCell}{Output}
  x^2/(y (-1 + x + y)) - x^2/(y (x + y))
  
\end{mmaCell}
\begin{mmaCell}{Output}
  1/y - 2/(-1 + x + y) + 1/(y (-1 + x + y)) + 
  y/(-1 + x + y) - y/(x + y)
  
\end{mmaCell}

The {\tt "IterativeGroebner"} option controls whether the polynomial reduction is performed iteratively (one inverse denominator factor at a time) or in a single step. The iterative approach is enabled by default to match {\tt MultivariateApart} and usually avoids intermediate expression swell; on small inputs, however, the single-step reduction can be the faster of the two, as the example below illustrates.
\begin{mmaCell}{Input}
  \mmaDef{expr} = 1/(y (x + y)(x + y - 1));

  \mmaDef{tmp1}=LinApart[\mmaDef{expr}, {x, y},
    "Method" -> "Groebner"];//AbsoluteTiming
  \mmaDef{tmp2}=LinApart[\mmaDef{expr}, {x, y}, 
    "Method" -> "Groebner",
    "IterativeGroebner" -> False];//AbsoluteTiming
  
  \mmaDef{tmp2}-\mmaDef{tmp1}
  
\end{mmaCell}
\begin{mmaCell}{Output}
  \{0.007566, Null\}
  \{0.003358, Null\}

  0
  
\end{mmaCell}

\subsection{Parallelization}

In the case of the method {\tt "MultivariateResidue"} the residue contributions from different bases are independent and can be computed simultaneously. The {\tt "Parallel"} option enables this parallelization for both univariate and multivariate decomposition,
\begin{mmaCell}{Input}
\mmaDef{LaunchKernels}[6]
\mmaDef{\$TemporaryDirectory}=NotebookDirectory[]

\mmaDef{expr}=1/(x y z (x+y)^2(y+z-1)^3(x+z-2)^5(x+y+z-3)^3);
\mmaDef{tmp1}=LinApart[\mmaDef{expr}, \{x, y, z\}];//AbsoluteTiming
\mmaDef{tmp2}=LinApart[\mmaDef{expr}, \{x, y, z\},
    "Parallel" -> \{True, 4, \mmaDef{\$TemporaryDirectory}\}];//AbsoluteTiming

\mmaDef{tmp1}-\mmaDef{tmp2}

\mmaDef{CloseKernels[]}
    
\end{mmaCell}
\begin{mmaCell}{Output}
  \{14.9985, Null\}
  \{19.7148, Null\}
  
  0
  
\end{mmaCell}
The {\tt "Parallel"} option takes a list of three arguments: a boolean flag, the number of cores, and a path for temporary files.

However, the effectiveness of such a routine is highly language dependent, as each system treats sub-workers (or subkernels) and memory management differently. In {\sc Wolfram Mathematica}, parallelization regarding symbolic computations is notoriously challenging\footnote{For parallelization we have followed the strategy outlined in the blog post \url{https://community.wolfram.com/groups/-/m/t/3540598?p_p_auth=zb1K16fL}.}. In our experience {\sc Mathematica}'s memory management limits the speed-up obtainable this way. The example above even shows a net overhead. We attribute this largely to {\sc Mathematica}'s memory model rather than to the algorithm itself, whose basis contributions are genuinely independent. In environments that allow finer control over memory we expect this overhead to be reduced.

\begin{table}[H]
\centering
\renewcommand{\arraystretch}{1.3}
\begin{tabular}{|p{0.32\textwidth}|p{0.58\textwidth}|}
\hline
\textbf{New Option} & \textbf{Description} \\
\hline\hline
{\tt Method} &
{ \footnotesize
Selects the multivariate decomposition method. Available choices: {\tt "MultivariateResidue"} (default for multivariate), {\tt "Leinartas"}, {\tt "Groebner"}. The univariate methods {\tt "ExtendedLaurentSeries"}, {\tt "Euclidean"}, and {\tt "EquationSystem"} from previous versions remain available.
}
\\
\hline
{\tt IterativeGroebner} &
{ \footnotesize
{\tt True}/{\tt False}: If set to {\tt True}, the Gr\"obner polynomial reduction is performed iteratively, one inverse denominator factor at a time, to avoid intermediate expression swell. Only relevant for {\tt "Method" -> "Groebner"}. Default: {\tt True}.
}
\\
\hline
\end{tabular}
\caption{\label{tab:new_options}
New options introduced in {\tt LinApart3}. The options from previous versions ({\tt Factor}, {\tt GaussianIntegers}, {\tt Extension}, {\tt Parallel}, {\tt PreCollect}, {\tt ApplyAfterPreCollect} and univariate choices for {\tt Method}) remain available and, excluding the last one, apply equally to the multivariate case.}
\end{table}

 
\section{Benchmarks}
\label{sect:benchmarks}
 
In this section, we compare the performance of four multivariate partial fraction decomposition implementations: the multivariate residue method ({\tt LinApart3}), the Leinartas method and the Gröbner basis method (both implemented in the {\tt LinApart} package) as well as the {\tt MultivariateApart} package~\cite{MultivariateApart}, on which we based our own Gröbner-method implementation.

We would like to emphasize that the algorithms presented have language-specific implementations. Their performance is strongly tied to the computer algebra environment in which they were written, thus we only comment on their scaling and relation to each other.

We would also like to note that we are not aware of any option in {\tt MultivariateApart} to restrict the decomposition to a subset of variables; consequently, {\tt MultivariateApart} always decomposes with respect to all variables and is excluded from the spectator-variable benchmark.
\\

During our benchmarks we measured the time and memory usage as a function of various complexity factors.

In general, a rational fraction's complexity can come from several factors, for example:
    \begin{enumerate}[label=(\roman*)]
        \item the number of distinct denominator factors;
        \item the number of variables;
        \item the algebraic complexity of the polynomial coefficients of the denominators;
        \item the multiplicity (the exponent) of the denominators;
        \item the degree of the numerator.
    \end{enumerate}

Since each algorithm exhibits different sensitivities to these factors, we varied one complexity parameter at a time (subject to resource constraints) to isolate their effects. We restricted ourselves to decomposition computations that were completed under 600s using less than 16 GB of RAM. We think these limits reflect practical constraints and typical real-world usage.
\\
 
\begin{figure}[!htbp]
\vspace*{-1cm}
    \begin{subfigure}[t]{\textwidth}
        \hspace*{-2.0cm}
        \centering
        \begin{minipage}{0.5\textwidth}
            \centering
            \includegraphics[scale=1.5]{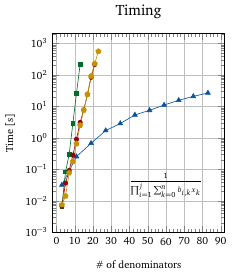}
        \end{minipage}%
        \begin{minipage}{0.5\textwidth}
            \centering
            \includegraphics[scale=1.5]{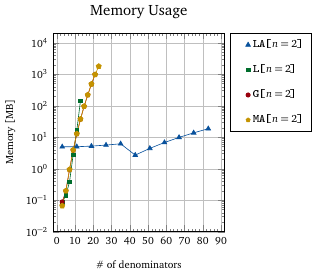}
        \end{minipage}
        \caption{\label{subfig:plot_for_num_of_denoms_N2} Benchmark for increasing number of linearly independent denominators with multiplicity one and two variables.}
    \end{subfigure}

    \par\vspace{1.5cm}     

    \begin{subfigure}[t]{\textwidth}
        \hspace*{-2.0cm}
        \centering
        \begin{minipage}{0.5\textwidth}
            \centering
            \includegraphics[scale=1.5]{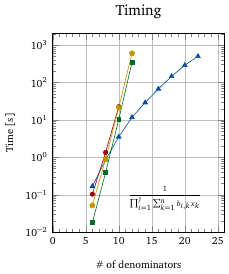}
        \end{minipage}%
        \begin{minipage}{0.5\textwidth}
            \centering
            \includegraphics[scale=1.5]{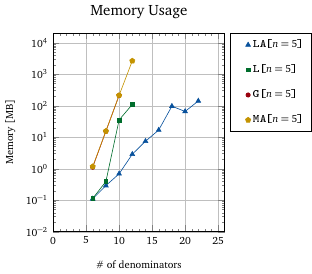}
        \end{minipage}
        \caption{\label{subfig:plot_for_num_of_denoms_N5} Same as Figure~\ref{subfig:plot_for_num_of_denoms_N2} but with multiplicity one and five variables.}
    \end{subfigure}
    
\caption{\label{fig:Benchmarks_NumOfDenoms1} Timings and memory usage of the new {\tt LinApart} function, our own implementation of the Leinartas method, the Gr\"obner basis method and {\tt MultivariateApart} (denoted as \texttt{LA}, \texttt{L}, \texttt{G} and \texttt{MA} in the legend) in case of different rational functions with numeric polynomial coefficients. In Figure~\ref{subfig:plot_for_num_of_denoms_N2} we plotted the benchmarks with increasing number of denominators with two variables ($n=2$), while in Figure~\ref{subfig:plot_for_num_of_denoms_N5} we show the same metrics for five variables ($n=5$).}

\end{figure}

We first investigated how the runtime scales with the number of distinct denominator factors ($j$), keeping the number of decomposition variables ($n$) fixed:
    \begin{gather}
    \label{function_for_num_of_denoms}
    f(x) 
    =
    \frac{1}{\prod_{i=1}^{j} \sum_{k=0}^{n} b_{i,k} x_k}
    \,,
    \end{gather}
where $x_0 \equiv 1$, the constant term.
The denominators were chosen to be generically linearly independent, with unit multiplicities and a constant numerator equal to one, while constants ($b_{i,k}$) were random integers from the interval $[10^4, 10^5]$; we considered two cases: $n = 2$ and $n = 5$. Our results can be seen on Figure~\ref{fig:Benchmarks_NumOfDenoms1}.

Over the range tested, the residue method scales markedly better in both cases: its runtime grows roughly polynomial while the other methods grow much faster, consistent with exponential behavior. As a result it can be orders of magnitude faster than the other tested methods on these inputs, with the advantage setting in later as the number of variables grows.

\begin{figure}[!htbp]
\vspace*{-1cm}
    \begin{subfigure}[t]{\textwidth}
        \hspace*{-2.0cm}
        \centering
        \begin{minipage}{0.5\textwidth}
            \centering
            \includegraphics[scale=1.5]{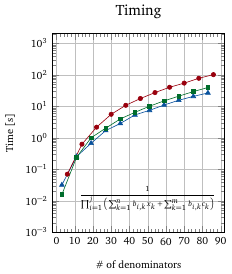}
        \end{minipage}%
        \begin{minipage}{0.5\textwidth}
            \centering
            \includegraphics[scale=1.5]{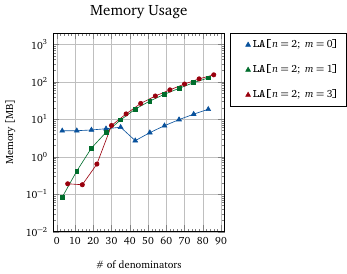}
        \end{minipage}
        \caption{\label{subfig:plot_for_num_of_denoms_residue_N2MX} Benchmark with the residue method for increasing number of linearly independent denominators with multiplicity one, two partial fraction variables and different number of spectator variables.}
    \end{subfigure}

    \par\vspace{1.5cm}     

    \begin{subfigure}[t]{\textwidth}
        \hspace*{-2.0cm}
        \centering
        \begin{minipage}{0.5\textwidth}
            \centering
            \includegraphics[scale=1.5]{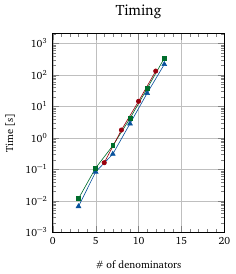}
        \end{minipage}%
        \begin{minipage}{0.5\textwidth}
            \centering
            \includegraphics[scale=1.5]{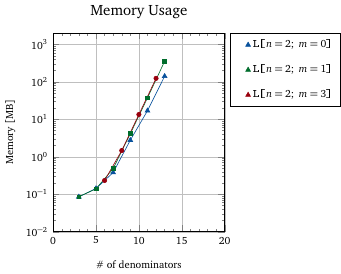}
        \end{minipage}
        \caption{\label{subfig:plot_for_num_of_denoms_leinartas_N2MX} Same as Figure~\ref{subfig:plot_for_num_of_denoms_residue_N2MX} but with the Leinartas method.}
    \end{subfigure}
    
\caption{\label{fig:Benchmarks_NumOfDenoms2} Timings and memory usage of the new {\tt LinApart} function (a) and our own implementation of the Leinartas method (b) in case of different rational functions with numeric polynomial coefficients, two partial fraction variables and different number of spectator variables.}

\end{figure}

\begin{figure}[!htbp]
\vspace*{-1cm}

    \begin{subfigure}[t]{\textwidth}
        \hspace*{-2.0cm}
        \centering
        \begin{minipage}{0.5\textwidth}
            \centering
            \includegraphics[scale=1.5]{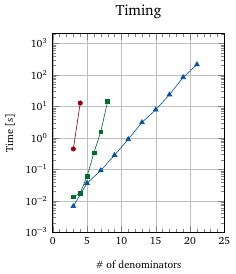}
        \end{minipage}%
        \begin{minipage}{0.5\textwidth}
            \centering
            \includegraphics[scale=1.5]{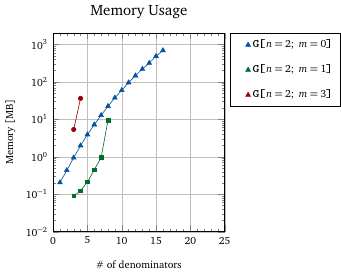}
        \end{minipage}
    \end{subfigure}
    
\caption{\label{fig:Benchmarks_NumOfDenoms3} Same as Figure~\ref{fig:Benchmarks_NumOfDenoms2} but with the Gr\"obner-basis method.}

\end{figure}

The Gröbner method highly depends on the number of spectator variables, since those complicate the polynomial ring causing massive intermediate swell during the Gröbner-basis calculation. In order to investigate the differences between the different methods in this regard, we generated rational functions with Eq.~\ref{function_for_num_of_denoms} but only decomposed in two variables. Our results are plotted on Figure~\ref{fig:Benchmarks_NumOfDenoms2} and Figure~\ref{fig:Benchmarks_NumOfDenoms3}.

As one can deduce from Figure~\ref{fig:Benchmarks_NumOfDenoms2} 
both the residue and Leinartas methods are insensitive to the number of spectator variables, whereas Figure~\ref{fig:Benchmarks_NumOfDenoms3} shows a steep dependence for the Gr\"obner method that effectively rules out its use when spectator variables are present (e.g. loop integrals with many masses).
\\

\begin{figure}[!htbp]
\vspace*{-1cm}
    \begin{subfigure}[t]{\textwidth}
        \hspace*{-2.0cm}
        \centering
        \begin{minipage}{0.5\textwidth}
            \centering
            \includegraphics[scale=1.5]{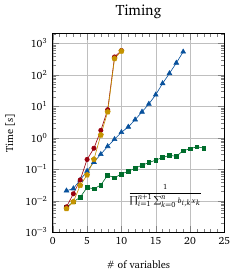}
        \end{minipage}%
        \begin{minipage}{0.5\textwidth}
            \centering
            \includegraphics[scale=1.5]{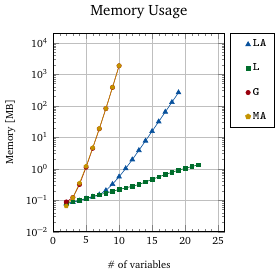}
        \end{minipage}
        \caption{\label{subfig:plot_for_num_of_vars_K1} Benchmark for increasing number of partial fraction variables with one linearly independent auxiliary denominator.}
    \end{subfigure}

    \par\vspace{1.5cm}     

    \begin{subfigure}[t]{\textwidth}
        \hspace*{-2.0cm}
        \centering
        \begin{minipage}{0.5\textwidth}
            \centering
            \includegraphics[scale=1.5]{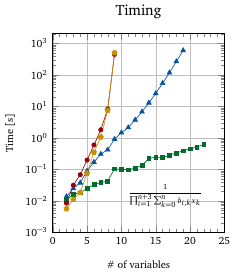}
        \end{minipage}%
        \begin{minipage}{0.5\textwidth}
            \centering
            \includegraphics[scale=1.5]{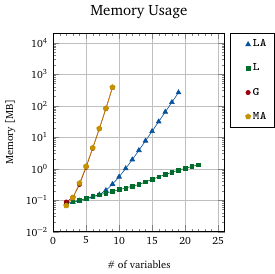}
        \end{minipage}
        \caption{\label{subfig:plot_for_num_of_vars_K3} Same as Figure~\ref{subfig:plot_for_num_of_vars_K1} but with three linearly independent auxiliary denominators.}
    \end{subfigure}
    
\caption{\label{fig:Benchmarks_NumOfVars} Timings and memory usage of the new {\tt LinApart} function, our own implementation of the Leinartas method, the Gr\"obner basis method and {\tt MultivariateApart} (denoted as \texttt{LA}, \texttt{L}, \texttt{G} and \texttt{MA} in the legend) in case of different rational functions with numeric polynomial coefficients. In Figure \ref{subfig:plot_for_num_of_vars_K1} and Figure \ref{subfig:plot_for_num_of_vars_K3} we plotted the benchmarks with increasing number of partial fraction variables with one and three linearly independent auxiliary denominators.}

\end{figure}

Next we investigated the limit of the high variable number case. In order to benchmark this case we constructed such function that had linearly independent denominators, $n$ number of variables and $a=1,3$ number of auxiliary denominators; the polynomial constants ($b_{i,k}$) were random integers from the interval $[10^4, 10^5]$
    \begin{gather}
    f(x) 
   =
    \frac{1}{\prod_{i=1}^{n+a} \sum_{k=0}^{n} b_{i,k} x_k}.
    \end{gather}
The timings and memory usage comparison can be seen on Figure~\ref{fig:Benchmarks_NumOfVars}.

As one could deduce from the graphs the implementations based on the Gröbner method performed the worst, while the Leinartas algorithm was the best and the residue method scaled somewhere in the middle. The performance drop of the Gröbner method stems from its reliance on Gröbner-basis algorithms, which have the tendency to significantly slow down as the polynomial ring grows in complexity (as seen before). The Leinartas algorithm shows the best scaling here; this is expected since it is only looking for affine relations between the denominators and then iteratively applies them. Hence the low auxiliary cases are best case scenarios for this algorithm, because increasing the number of variables only raises the rank of the matrix it has to invert, not the number of iterations, which are the main bottleneck of this method. For the residue method, more partial-fraction variables yield increasingly many bases, thus we have to calculate significantly more residues.
\\

\begin{figure}[!htbp]
\vspace*{-1cm}
    \begin{subfigure}[t]{\textwidth}
        \hspace*{-2.0cm}
        \centering
        \begin{minipage}{0.5\textwidth}
            \centering
            \includegraphics[scale=1.5]{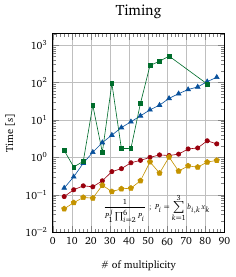}
        \end{minipage}%
        \begin{minipage}{0.5\textwidth}
            \centering
            \includegraphics[scale=1.5]{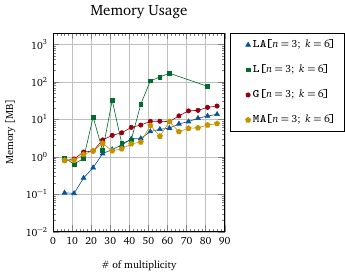}
        \end{minipage}
        \caption{\label{subfig:plot_for_multiplicity_one_N3K6} Benchmark for increasing the multiplicity of one denominator of a fraction with linearly independent denominators with $n=3$ variables and $k=6$ denominators, whose polynomial coefficients are ``high'' random integer numbers.}
    \end{subfigure}

    \par\vspace{1.5cm}     

    \begin{subfigure}[t]{\textwidth}
        \hspace*{-2.0cm}
        \centering
        \begin{minipage}{0.5\textwidth}
            \centering
            \includegraphics[scale=1.5]{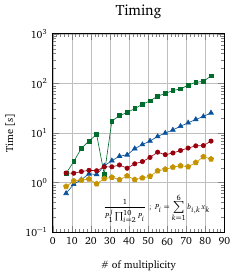}
        \end{minipage}%
        \begin{minipage}{0.5\textwidth}
            \centering
            \includegraphics[scale=1.5]{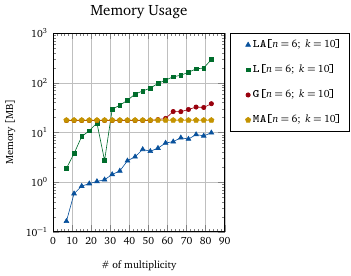}
        \end{minipage}
        \caption{\label{subfig:plot_for_multiplicity_one_N6K10} The same as Figure~\ref{subfig:plot_for_multiplicity_one_N3K6} but with $n=6$ variables and $k=10$.}
    \end{subfigure}
    
\caption{\label{fig:Benchmarks_Multiplicity_One} Timings and memory usage of the new {\tt LinApart} function, our own implementation of the Leinartas method, the Gr\"obner basis method and {\tt MultivariateApart} (denoted as \texttt{LA}, \texttt{L}, \texttt{G} and \texttt{MA} in the legend) in cases of increasing multiplicity of one denominator of a fraction with linearly independent denominators with different variable number and denominator number, where polynomial coefficients were ``high'' random integer numbers.}

\end{figure}

One of the most interesting cases is that of the higher multiplicities, since in our previous articles the univariate algorithms were highly sensitive to such increases. We first investigated the case, when one denominator possessed an increasingly higher multiplicity ($j$):
    \begin{gather}
    f(x) 
   =
    \frac{1}{P_1^j \prod_{i=2}^{k} P_i},~\text{where}~P_i=\sum_{l=0}^{n} b_{i,l} x_l
    \end{gather}
and $n=3,6$, $k=6,10$, furthermore the coefficients ($b_{i,l}$) were random integers from the range $[10^4, 10^5]$. Our results are plotted on Figure~\ref{fig:Benchmarks_Multiplicity_One}.

The Figure~\ref{subfig:plot_for_multiplicity_one_N3K6} shows a highly fluctuating number for the Leinartas method, which we got in low variable cases but not in high variable cases. Under further investigation we concluded that this is a heuristic effect, connected to the complexity of finding affine relations.

In both cases the Gröbner-basis approach is the most efficient. Comparing Figure~\ref{subfig:plot_for_multiplicity_one_N3K6} and Figure~\ref{subfig:plot_for_multiplicity_one_N6K10}, one sees the effect of the different variable number: the gap between the best (Gröbner-basis) and second best (residue) shrinks markedly.

The reason for the bad scaling of the Leinartas method is due to its iterative nature; the higher the multiplicity, the higher the number of iterations and expansions, which it must perform. 

For the residue method, high multiplicities cause a swell in the number of terms, hence the steep scaling.

This case is highly favorable for the Gröbner-basis method, since the multiplicity only influences the last step of the algorithm, namely the polynomial reduction. Furthermore, since we only increased one multiplicity during the reduction the algorithm only has to do one high power expansion, hence the low sensitivity.

\begin{figure}[!htbp]
\vspace*{-1cm}
    \begin{subfigure}[t]{\textwidth}
        \hspace*{-2.0cm}
        \centering
        \begin{minipage}{0.5\textwidth}
            \centering
            \includegraphics[scale=1.5]{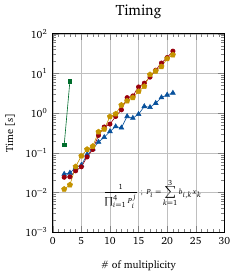}
        \end{minipage}%
        \begin{minipage}{0.5\textwidth}
            \centering
            \includegraphics[scale=1.5]{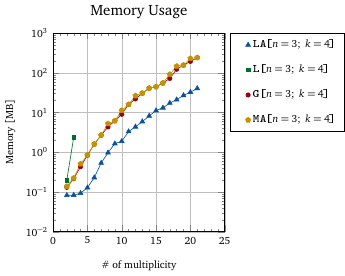}
        \end{minipage}
        \caption{\label{subfig:plot_for_multiplicity_many_N3K4} Benchmark for increasing multiplicity of all denominators of a fraction with linearly independent denominators with $n=3$ variables and $k=4$ denominators, where polynomial coefficients are ``high'' random integer numbers.}
    \end{subfigure}

    \par\vspace{1.5cm}     

    \begin{subfigure}[t]{\textwidth}
        \hspace*{-2.0cm}
        \centering
        \begin{minipage}{0.5\textwidth}
            \centering
            \includegraphics[scale=1.5]{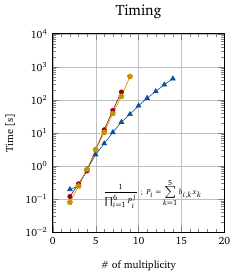}
        \end{minipage}%
        \begin{minipage}{0.5\textwidth}
            \centering
            \includegraphics[scale=1.5]{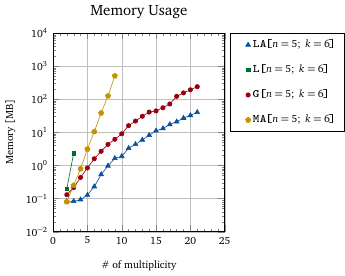}
        \end{minipage}
        \caption{\label{subfig:plot_for_multiplicity_many_N56} Same as Figure~\ref{subfig:plot_for_multiplicity_many_N3K4} but with $n=5$ and $k=6$.}
    \end{subfigure}
    
\caption{\label{fig:Benchmarks_Multiplicity_Many} Timings and memory usage of the new {\tt LinApart} function, our own implementation of the Leinartas method, the Gr\"obner basis method and {\tt MultivariateApart} (denoted as \texttt{LA}, \texttt{L}, \texttt{G} and \texttt{MA} in the legend) in cases of increasing multiplicity of all denominators of a fraction with linearly independent denominators with different variable number and denominator number, whose polynomial coefficients were ``high'' random integer numbers.}

\end{figure}

It is therefore also important to investigate the case, when all of the denominators' multiplicities are increased ($j$):
    \begin{gather}
    f(x) 
   =
    \frac{1}{ \prod_{i=1}^{k} P_i^j},~\text{where}~P_i=\sum_{l=0}^{n} b_{i,l} x_l
    \end{gather}
and $n=3,5$, $k=4,6$ and the coefficients ($b_{i,l}$) were random integers from the range $[10^4, 10^5]$. The results are shown on Figure~\ref{fig:Benchmarks_Multiplicity_Many}.

In this case during the Gröbner-basis method the polynomial reduction becomes exponential over the tested ranges, since it has to expand more expressions with high power, while the residue method scales better and similarly to the previous case, because only the number of bases increased. The iterative method, by contrast, scales very poorly here: with our implementation it could not even decompose ``basic'' fractions, which posed no problem to the other two approaches.
\\

\begin{figure}[!htbp]
\vspace*{-1cm}
    \begin{subfigure}[t]{\textwidth}
        \hspace*{-2.0cm}
        \centering
        \begin{minipage}{0.5\textwidth}
            \centering
            \includegraphics[scale=1.5]{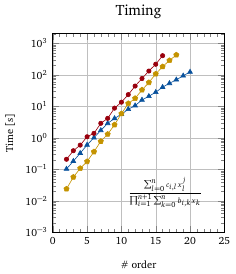}
        \end{minipage}%
        \begin{minipage}{0.5\textwidth}
            \centering
            \includegraphics[scale=1.5]{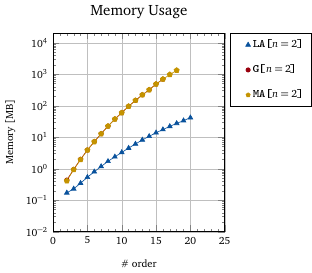}
        \end{minipage}
        \caption{\label{subfig:plot_for_order_of_numerator_N2K1} Benchmark for increasing order of the numerator with two variables and three denominators, with ``high'' random integer numbers as polynomial coefficients.}
    \end{subfigure}

    \par\vspace{1.5cm}     

    \begin{subfigure}[t]{\textwidth}
        \hspace*{-2.0cm}
        \centering
        \begin{minipage}{0.5\textwidth}
            \centering
            \includegraphics[scale=1.5]{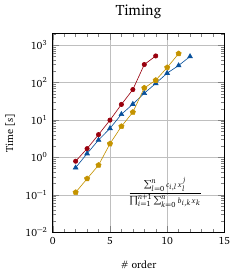}
        \end{minipage}%
        \begin{minipage}{0.5\textwidth}
            \centering
            \includegraphics[scale=1.5]{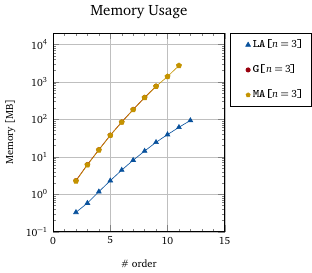}
        \end{minipage}
        \caption{\label{subfig:plot_for_order_of_numerator_N3K1} Same as Figure~\ref{subfig:plot_for_order_of_numerator_N2K1} but with three variables and four denominators.}
    \end{subfigure}
    
\caption{\label{fig:Benchmarks_OrderOfNumerator} Timings and memory usage of the new {\tt LinApart} function, our own implementation of the Gr\"obner basis method and {\tt MultivariateApart} (denoted as \texttt{LA}, \texttt{G} and \texttt{MA} in the legend) in cases of increasing order of the numerator with two or three variables and four denominators, with ``high'' random integer numbers as polynomial coefficients.}

\end{figure}

Another factor that influences the polynomial reduction, and hence the run-time, is the order of the numerator ($j$). To test this case we used a fraction like,
    \begin{gather}
    f(x) 
   =\frac{\sum_{l=0}^{n} c_{l} x_l^j}
            {\prod_{i=1}^{n+1} \sum_{k=0}^{n} b_{i,k} x_k},
    \end{gather}
where the constants ($c_{l}$ and $b_{i,k}$) were random integers from the range $[10^4, 10^5]$ and $n$ took the values $n=2,3$. Our findings are summarized on Figure~\ref{fig:Benchmarks_OrderOfNumerator}.

To the best of our knowledge the Leinartas method does not contain any steps regarding non-unity numerators, so we have excluded it from this benchmark. Based on the curves of Figure~\ref{fig:Benchmarks_OrderOfNumerator} one can conclude that {\tt LinApart3} with its numerator reduction step possesses a better scaling than the Gröbner-basis method.
\\

\begin{figure}[!htbp]
\vspace*{-1cm}
    \begin{subfigure}[t]{\textwidth}
        \hspace*{-2.0cm}
        \centering
        \begin{minipage}{0.5\textwidth}
            \centering
            \includegraphics[scale=1.5]{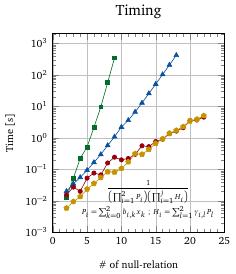}
        \end{minipage}%
        \begin{minipage}{0.5\textwidth}
            \centering
            \includegraphics[scale=1.5]{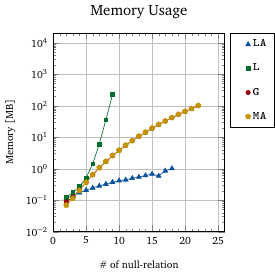}
        \end{minipage}
        \caption{\label{subfig:plot_for_nullrel_M2} Benchmark for increasing order of null-relations between the denominators in case of two variables and two linearly independent denominators.}
    \end{subfigure}

    \par\vspace{1.5cm}     

    \begin{subfigure}[t]{\textwidth}
        \hspace*{-2.0cm}
        \centering
        \begin{minipage}{0.5\textwidth}
            \centering
            \includegraphics[scale=1.5]{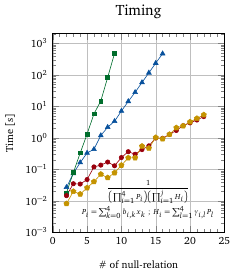}
        \end{minipage}%
        \begin{minipage}{0.5\textwidth}
            \centering
            \includegraphics[scale=1.5]{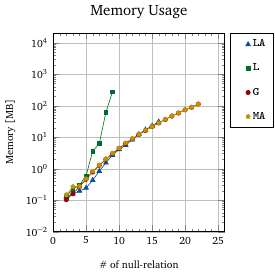}
        \end{minipage}
        \caption{\label{subfig:plot_for_nullrel_M4} Same as Figure~\ref{subfig:plot_for_nullrel_M2} but for four variables and four linearly independent denominators.}
    \end{subfigure}
    
\caption{\label{fig:Benchmarks_NullRel} Timings and memory usage of the new {\tt LinApart} function, our own implementation of the Leinartas method, the Gr\"obner basis method and {\tt MultivariateApart} (denoted as \texttt{LA}, \texttt{L}, \texttt{G} and \texttt{MA} in the legend) in cases of increasing number of null-relations between the denominators.}

\end{figure}

Up to this point we only considered fractions without any null-relations; however, in real-life use-cases this is not always true. For this purpose we have generated fractions in the form:
    \begin{gather}
    \label{eq_with_null_rellation}
    f(x) 
   =\dfrac{1}{
    	    \Big( \prod_{i=1}^{n} P_i \Big)
	        \Big( \prod_{i=1}^{j} H_i \Big)
    	  },~\text{where}~P_i=\sum_{k=0}^{n} b_{i,k} x_k~\text{and}~H_i = \sum_l \gamma_{i,l} P_l.
    \end{gather}
In this case we increased the variable count $j$, set $n=3,4$ and the constants ($b_{i,k}$) to random integers from the range $[10^4,10^5]$. Timings and memory consumptions can be seen on Figure~\ref{fig:Benchmarks_NullRel}.

Based on the curves our assumptions were confirmed, namely the approach based on the Gröbner-basis has superior scaling due to the fact that it basically does the null-relation elimination and the ideal finding in one step. Thus it only does one time-consuming step, while the residue method first does the elimination and only then calculates the residues. In this case the Leinartas algorithm has the worst scaling, since it has to iteratively get rid of the dependencies.

\subsection{Summary of benchmarks}
\label{subsect:bench_summary}

We have summarized our qualitative findings in Table~\ref{tab:benchmark_summary}. As said before, the exact timings can be implementation and programming language dependent. So, we focus on qualitative scaling trends rather than absolute timings. Based on these trends we believe the residue method is the most promising all-around algorithm for multivariate partial fraction  decomposition of expressions with linear denominators. Its favorable, apparently polynomial growth with the number of denominators, its insensitivity to spectator variables, and its mild dependence on the number of decomposition variables make it particularly well-suited for the expressions arising in perturbative quantum field theory, where tens of denominators with symbolic coefficients and few or no null relations are common. Moreover, unlike {\tt MultivariateApart}, it allows the user to choose which variables to decompose without introducing spurious singularities.

\begin{table}[H]
\centering
\hspace*{-1.5cm}
\renewcommand{\arraystretch}{1.3}
\begin{tabular}{|l|c|c|c|c|}
\hline
\textbf{Parameter} & \textbf{Residue} & \textbf{Leinartas}
  & \textbf{Gr\"obner} & \textbf{MultivariateApart} \\
\hline\hline
\# denominators ($k$)
  & polynomial & exponential & exponential & exponential \\
\hline
Spectator variables
  & insensitive & insensitive & severe & N/A$^*$ \\
\hline
\# variables ($n$)
  & mild & mild & severe & severe \\
\hline
Multiplicities of one denominator
  & mild & severe & moderate & moderate \\
\hline
Multiplicities of many denominators
  & mild & severe & exponential & exponential \\
\hline
Numerator degree
  & moderate & N/A$^\ddagger$ & mild & mild \\
\hline
Null relations
  & severe & severe & mild & mild \\
\hline
\end{tabular}
\caption{\label{tab:benchmark_summary}
Summary of scaling behavior we observed in the tested ranges. $^*${\tt MultivariateApart}
  cannot restrict the decomposition to a subset of variables. $^\ddagger$The Leinartas decomposition does not
  reduce numerator degree.}
\end{table}

\section{FORM implementation}
\label{sect:FORM}

In addition to the {\sc Wolfram Mathematica} package described in
Section~\ref{sect:Mathematica}, we provide an implementation of
{\tt LinApart3} in the \texttt{FORM} computer algebra
system~\cite{Davies:2026cci}.  The motivation is twofold.  First,
expressions arising in perturbative QCD, such as amplitude residues,
IBP master-integral coefficients, and integrands prepared for
parametric integration, routinely outgrow what {\sc Mathematica}
can process efficiently.  \texttt{FORM}, on the other hand, is built for
handling large symbolic expressions in bulk and runs well on many
cores.  Second, an in-system partial fraction routine fits
naturally into existing \texttt{FORM}-based pipelines for analytic
integration, for example
\texttt{HyperFORM}~\cite{Kardos:2025klp}, where the rational
functions are produced and used by \texttt{FORM} itself.  Running
the partial-fraction step inside \texttt{FORM} avoids sending each
expression to another computer algebra system.

Throughout this section we refer repeatedly to \emph{Phase~1} and
\emph{Phase~2} of the algorithm, in the sense established in
Section~\ref{sect:algorithm}: Phase~1 is the null-relation
elimination step, which uses linear relations among the
denominators to reduce every term until at most $n$ distinct
denominators remain.  Phase~2 is the basis-identification and
residue-extraction step, which then applies the multivariate
Grothendieck residue formula to the surviving terms.  The two
phases are implemented as separate sub-pipelines and discussed
separately below.

\subsection{Installation}
\label{subsect:FORMinstall}

The \texttt{FORM} implementation of {\tt LinApart3} consists of
two source files.  The first, \texttt{declare-formapart.h},
declares the symbols, functions, tables, and preprocessor
variables used by the library and sets the shipped defaults for
the user-tunable preprocessor flags \texttt{APglobalOrder},
\texttt{APorder\BR Descending}, \texttt{APuseBareiss}, and
\texttt{APuse\BR Global\BR Null\BR Relations}, all four set to \texttt{"0"} by default.
The second, \texttt{formapart.h}, defines the procedures that
implement the steps of Section~\ref{sect:algorithm} and pulls in
the declarations automatically.  Standalone test, regression, and
benchmark \texttt{.frm} files are also shipped with the package.  A
user script only needs to include \texttt{formapart.h} near the top:
\begin{lstlisting}
#include formapart.h
\end{lstlisting}
The procedures from the library are then called through the usual
\texttt{FORM} \texttt{\#call} mechanism (see
Section~\ref{subsect:FORMusage} for a worked example).

We recommend adding the directory with the two sources to the
\texttt{FORMPATH} environment variable, so that \texttt{FORM}'s
preprocessor finds the include without an absolute path.  This keeps user scripts
portable across machines and install locations.  In a Bourne-style
shell, add to the startup file:
\begin{lstlisting}[language={}]
export FORMPATH="/path/to/LinApart/form:$FORMPATH"
\end{lstlisting}
with \texttt{/path/to/LinApart/form} replaced by the actual install
location.

The library requires \texttt{FORM5}~\cite{Davies:2026cci}.
Specifically, we rely on the relaxed placement rules for the
\texttt{ModuleOption} statement introduced there (Section 2.7 of
ref.~\cite{Davies:2026cci}).  This is what enables the per-term
state design described in Section~\ref{subsect:FORMalgos}.  Under
earlier \texttt{FORM} releases the same declarations would have to
be lifted to the end of each module, which is more inconvenient
for a procedure library of this shape.

The \texttt{FORM} sources live in the same repository as the
{\sc Mathematica} package at
\url{https://github.com/fekeshazy/LinApart}, under
\texttt{form/}.  A Ruby-based unit-test driver
(\texttt{check.rb}, originally part of the \texttt{FORM}
distribution) is included: it reads the multi-fold \texttt{.frm}
test files, splits each into its individual folds, feeds every
fold to \texttt{FORM}, and evaluates Ruby-side assertions against
the captured output.  No build step is required.

\subsection{Implementation strategy}
\label{subsect:FORMalgos}

The \texttt{FORM} implementation is not a straight translation of
the {\sc Mathematica} routines of Section~\ref{sect:Mathematica}.
The two systems work very differently: {\sc Mathematica} operates
on a global expression tree with random-access mutation and lookup,
while \texttt{FORM} processes expressions as streams of terms: the
terms only stream through the current module, a \texttt{.sort}
statement then sorts the expression(s), and the stream starts again
with the possibly new set of terms.  The
fixed data the algorithm needs lives in global ctables that are
built once during setup.  The denominator coefficient matrix and
the denominator polynomial table are present in every run, the
global denominator-elimination order is built when the
fill-reducing ordering is requested, and the null-relation cache
is built when the cover-first cache branch is selected.  The
per-term bookkeeping needed by the rewriter lives in
\emph{per-term} local dollar variables, set from each term's
structure as it passes through the module.  Examples include the
index of a denominator about to be eliminated, the matrix entries
of an adjugate cell, and swap indices for pivoting.  The
implementation is built around this split.

\paragraph{Per-term state via dollars}
The pattern is to seed a dollar from the term's structure, refine
it under \texttt{Inside}/\texttt{EndInside}, and then declare
\texttt{ModuleOption,Local} so that each \texttt{FORM} worker has
its own copy of the dollar within the current module.  An excerpt
from \texttt{Apart\BR Update\BR Active\BR Denominators}, which looks up an
eliminated denominator's index in the global denominator pool so
the corresponding bit in the active-set collector can be flipped,
makes the three steps concrete:
\begin{lstlisting}
$Position = `AuxFuncID'($TMPden);
Inside $Position;
  id `AuxFuncID'(`Sym1'?`DenSetID'[`Sym2']) = `Sym2';
EndInside;
ModuleOption,Local,$Position;
\end{lstlisting}

\paragraph{Mostly read-only ctables}
The denominator coefficient matrix \texttt{APcoeffTbl}, the
denominator polynomial table \texttt{APdenTbl}, and (when the
fill-reducing ordering is enabled) the elimination permutation
\texttt{APpermTbl} are filled once during setup and never written
to afterwards.  Phase~1 and Phase~2 only read from
them.  The null-relation cache \texttt{APnullTbl} is the one
exception: it is created \texttt{sparse} at the start of Phase~1
and grows during the per-term work (see the
\emph{cover-first cache} branch below).  The entire cache
derivation step is run single-threaded via a
\texttt{NotInParallel} declaration on the temporary expression
that holds the unique active-set configurations of the current
terms.  This covers both the table reads (cover walks back to
populated supersets) and the writes.  The per-term elimination
steps that follow run in parallel as usual. We decided to
run this step on a single core because profiling showed that
distributing the terms to workers outweighed the parallel speed-up for this step.

\paragraph{Two strategies for null-relation lookup}
For each input term we need the null relations applicable to its
active denominator set.  We offer two paths.  The \emph{determinant}
branch recomputes the relevant relations per term by Laplace
cofactor expansion of the active submatrices via the recursive
routine \texttt{Apart\BR Determinant\BR Thru\BR Cofactor\BR Expansion}.  This is
cheap when the term has few denominators but scales poorly when
many terms share large active sets.  The \emph{cover-first cache}
branch derives the null relations of the full set of denominators
once during setup and stores them under the maximal-bitmask key of
the sparse table \texttt{APnullTbl}.  Each Phase~1 elimination
iteration then collects the unique active-set bitmasks of the
current terms and, for every bitmask not yet in the table,
derives its relations from a cached superset by a Hamming-cover
walk (\texttt{Apart\BR Find\BR Cached\BR Cover}) followed by elimination of
the absent denominators
(\texttt{Apart\BR Remove\BR Absent\BR Denominators\BR From\BR Null\BR Relations}).
The new entries are written back into the table, and the per-term
elimination then fetches relations by direct bitmask lookup.  The
two branches are complementary.  We do not provide an automatic
selector, and recommend that users experiment with both flags on
representative inputs and pick whichever wins.
Section~\ref{subsect:FORMBenchmarks} gives examples of what such a
comparison looks like in practice.  A fraction-free Bareiss
elimination routine (\texttt{Apart\BR Bareiss\BR Determinant}) is also
provided as an alternative to plain cofactor expansion, selected
by the independent preprocessor flag \texttt{APuseBareiss}.  The
benchmarks below all run with the cofactor default.  The Bareiss
routine is most useful when the denominator coefficients are
numeric (integers or rationals).  In that setting the fraction-free
elimination keeps the intermediate expressions compact, whereas
with the symbolic coefficients typical of physics applications, the
recursive cofactor expansion is usually cheaper, since each
cofactor step only multiplies and adds entries, while the Bareiss
recurrence performs a polynomial division by the previous pivot at
every reduction step.

\paragraph{Fill-reducing denominator ordering}
The determinant path is sensitive to the order in which
denominators are presented: an ordering that minimizes fill-in during cofactor expansion is intended
to reduce intermediate expression sizes; this option is not exercised in the benchmarks below, which
use the default order.  Our parser optionally applies a fill-reducing
global ordering to each input before Phase~1.  The score used to
sort denominators is the number of null relations of the complete
denominator set in which a given denominator appears with a
non-zero coefficient
(\texttt{Apart\BR Build\BR Global\BR Denominator\BR Order}).  Denominators are
then sorted by this score (ascending by default, descending when
\texttt{APorder\BR Descending} is set); denominators with equal score
keep their original input order.  The ordering is computed once, before
Phase~1, and only determines which denominator is eliminated next, so
it applies unchanged whether Phase~1 uses the determinant branch or
the cover-first cache.

\paragraph{Dependency on \texttt{FORM5}}
The per-term dollar design depends on the relaxed
\texttt{Module\BR Option} placement of
\texttt{FORM5}~\cite{Davies:2026cci}, which allows
\texttt{ModuleOption,Local,\$x} declarations \emph{inside}
procedure bodies and inside nested preprocessor loops.  Under
earlier releases the same declarations had to be lifted to the end
of the module, with every dollar used anywhere in the module
listed there.  This is cumbersome for a procedure library where
dollar names are local to individual procedures and, in the
adjugate and matrix-inverse routines, are generated by the
preprocessor inside nested loops over runtime indices.  The
adjugate routine shows the pattern that is possible under
\texttt{FORM5}:
\begin{lstlisting}
#procedure ApartAdjugateMatrix(dMat,SepChar,
              MatrixDimension,RowFuncID,ColFuncID,
              AuxFuncID,TableID,Sym1,...,Sym4)
  id `RowFuncID'(?a)*`ColFuncID'(?b) =
    `AuxFuncID'(`RowFuncID'(?a)*`ColFuncID'(?b));
  id `AuxFuncID'(`Sym1'?$RowsAndColumns) = `Sym1';
  ModuleOption,Local,$RowsAndColumns;
  #Do iRow=1,`MatrixDimension'
    #Do jCol=1,`MatrixDimension'
      $`dMat'`iRow'`SepChar'`jCol' = $RowsAndColumns;
      Inside $`dMat'`iRow'`SepChar'`jCol';
        Multiply (-1)^(`iRow'+`jCol');
        ...
      EndInside;
      ModuleOption,Local,$`dMat'`iRow'`SepChar'`jCol';
    #EndDo
  #EndDo
#endprocedure
\end{lstlisting}
The dollar name is built at preprocess time from the caller-chosen
matrix-name stem \texttt{dMat}, the loop indices \texttt{iRow} and
\texttt{jCol}, and the separator character \texttt{SepChar}
(\texttt{SepChar} is not an index, but a single character glued
between the row and column indices to keep the resulting dollar
name unambiguous), and the \texttt{ModuleOption,Local} declaration
appears in the same scope.

Notice also the calling convention: every symbol, function head,
table, and set the routine touches is passed in as a named
argument (\texttt{RowFuncID}, \texttt{ColFuncID},
\texttt{AuxFuncID}, \texttt{TableID}, \texttt{Sym1, \dots, Sym4},
the matrix-name stem \texttt{dMat}, and the matrix-name separator
character \texttt{SepChar}) rather than hard-coded inside the
procedure body.  The same convention runs through the whole
library and is what makes the routines composable: a caller can
reuse \texttt{Apart\BR Adjugate\BR Matrix} on its own matrices and
auxiliary functions without colliding with the library's internal
naming, and the procedure itself can be invoked from several call
sites in a single \texttt{FORM} script with disjoint working
state.

\subsection{Usage}
\label{subsect:FORMusage}

A typical \texttt{FORM} script for this library is short.  We work the example of
Section~\ref{sect:example},
\begin{equation*}
  f(x, y) = \frac{1}{x^2\, y\, (x+y)\, (x+y-1)}\,,
\end{equation*}
so the \texttt{FORM} call can be compared line-by-line against the
{\sc Mathematica} invocation of Section~\ref{sect:Mathematica}.

\paragraph{Minimal script}
The complete \texttt{FORM} program that reproduces the partial-fraction
decomposition of this example is
\begin{lstlisting}
#include formapart.h
CFunction den;
Symbols x, y;
Local F = den(x)^2 * den(y) * den(x+y) * den(x+y-1);
.sort
#call ApartMultiLinApart(den,x,y)
Print +s;
.end
\end{lstlisting}
The denominator factors are wrapped in a single user-declared
\texttt{cfunction}, with multiplicities carried as integer
exponents on the wrapper.  The
wrapper's name (here \texttt{den}) is passed as the first argument
to \texttt{Apart\BR Multi\BR Lin\BR Apart} and may be chosen freely.  The
arguments of \texttt{den} must be linear in the variables that
follow it in the \texttt{\#call}, here \texttt{x} and \texttt{y}.
The parser does not check for linearity: it silently treats every
\texttt{den} argument as linear and extracts coefficients
accordingly via \texttt{div\_} and \texttt{rem\_}, so non-linear
arguments produce wrong output rather than an error.  Ensuring
that every \texttt{den} argument is linear in the active variables
is therefore the user's responsibility.  Spectator-only factors
are recognized by the parser
(\texttt{Apart\BR Tag\BR Var\BR Dep\BR Function\BR Heads}) and left untouched
whether they sit inside a \texttt{den} wrapper or not.  Any factor
not wrapped in \texttt{den} simply passes through the
decomposition unchanged.

\paragraph{Output}
Running the above through \texttt{FORM} prints
\begin{lstlisting}
F =
    + den(x)*den( - 1 + y + x)
    + den(x)^2*den( - 1 + y + x)
    - den(x)^2*den(y)
    - den(x)^2*den(y)^2
    - den(x)*den(y)
    + den(x)*den(y)^3
    + den(y)*den( - 1 + y + x)
    - den(y)^3*den(y + x)
   ;
\end{lstlisting}
Reading \texttt{den(a)} as $1/a$, this is
\begin{equation*}
\begin{aligned}
  f(x,y) &= \frac{1}{x(x{+}y{-}1)} + \frac{1}{x^2(x{+}y{-}1)}
         + \frac{1}{y(x{+}y{-}1)} \\
         &\quad - \frac{1}{x^2 y} - \frac{1}{x^2 y^2} - \frac{1}{x y}
         + \frac{1}{x y^3} - \frac{1}{y^3 (x{+}y)}\,.
\end{aligned}
\end{equation*}
This eight-term decomposition is algebraically equivalent to the
seven-term hand-worked result of
Section~\ref{sect:example}, eq.~\eqref{eq:example_result}, but
structurally different.  The difference comes
entirely from the priority ordering that decides which denominator
is eliminated at each step. As Section~\ref{sect:algorithm} notes,
this ordering is arbitrary as long as it is fixed before the recursion
begins, and the \texttt{FORM} and {\sc Mathematica} implementations
fix \emph{opposite} ones: \texttt{Apart\BR Multi\BR Lin\BR Apart}
eliminates the lowest-multiplicity member of each relation, whereas
{\sc Mathematica} eliminates the highest. For the relation
$D_1 + D_2 - D_3 = 0$ at hand this means \texttt{FORM} eliminates
$D_2 = y$, while {\sc Mathematica} and the hand-worked
Section~\ref{sect:example} eliminate $D_1 = x$. The two orderings
produce different but equivalent decompositions.  Both outputs satisfy the
guarantees of Section~\ref{sect:algorithm}: every term contains
at most $n=2$ distinct denominators drawn from the original set,
and no spurious singularities are introduced.

\paragraph{Active variables and spectators}
The active variables used for null elimination, basis
identification, and residue extraction are the variables passed to
\texttt{Apart\BR Multi\BR Lin\BR Apart} after the
\texttt{cfunction}-name argument.  Any other symbol
appearing in numerator or denominator coefficients, for example
kinematic invariants, masses, or the dimensional regulator
$\varepsilon$, is treated as a spectator parameter and passes
through the decomposition unchanged.  Spectator symbols need no
special declaration beyond the usual \texttt{FORM}
\texttt{symbol}/\texttt{symbols} statement.  This matches the
spectator-insensitivity reported for the {\sc Mathematica}
implementation (Section~\ref{subsect:bench_summary},
Table~\ref{tab:benchmark_summary}): neither implementation slows
down as the number of spectators grows.

\paragraph{Numeric round-trip verification}
We include a companion procedure \texttt{Apart\BR Numeric\BR Check} for
regression tests and one-off correctness checks.  It substitutes
large random primes for every active variable and spectator (a
Schwartz-Zippel-style witness) and reports the numeric difference
between the original and decomposed expressions.  A zero result is
strong evidence that the two are algebraically equal.  The usual
pattern is to copy the input before decomposing, hide the copy
while \texttt{Apart\BR Multi\BR Lin\BR Apart} runs, and compare afterwards:
\begin{lstlisting}
#include formapart.h
#define Vars   "x,y"
#define Params ""
CFunction den;
Symbols x, y;
Symbol  n1;
Local Finput  = den(x)^2 * den(y) * den(x+y) * den(x+y-1);
.sort
Local Foutput = Finput;
.sort
Hide Finput;
.sort
#call ApartMultiLinApart(den,x,y)
.sort
Unhide Finput;
.sort
#call ApartNumericCheck(NumDiff,Finput,Foutput,den,Vars,Params,n1)
.sort
Print +s NumDiff;
.end
\end{lstlisting}
The \texttt{Hide}/\texttt{Unhide} bracketing is necessary because
\texttt{Apart\BR Multi\BR Lin\BR Apart} acts on every visible local expression.
Hiding \texttt{Finput} during the call ensures only \texttt{Foutput}
is decomposed, and unhiding before the check restores the original
for comparison.  Successful checks print \texttt{NumDiff = 0;}.

\paragraph{Mode selection}
The two Phase 1 paths discussed in Section~\ref{subsect:FORMalgos}
are selected by preprocessor flags set immediately after the
\texttt{\#include} line:
\begin{lstlisting}
* Determinant branch:
#Redefine APuseGlobalNullRelations "0"
* Cover-first cache branch:
#Redefine APuseGlobalNullRelations "1"
\end{lstlisting}
The fill-reducing denominator ordering is controlled analogously:
\begin{lstlisting}
* 1 = enable fill-reducing ordering, 0 = disable:
#Redefine APglobalOrder     "1"
* 0 = ascending in the score, 1 = descending:
#Redefine APorderDescending "0"
\end{lstlisting}

On the realistic benchmark of
Section~\ref{subsect:FORMBenchmarks} the cache branch is the
more economic of the two in aggregate, so we suggest it as
the default starting point; as noted above we
provide no automatic selector, and the determinant branch is
worth trying on atypical inputs.

\subsection{Benchmarks}
\label{subsect:FORMBenchmarks}

To test our \texttt{FORM} implementation on a realistic and
publicly available example we benchmark it on the two-loop
non-planar five-point double-pentagon dlog-basis IBP coefficient
matrix of Ref.~\cite{Bendle:2019csk}: a $26 \times 108$ matrix of
rational functions of $\{\varepsilon, s_{15}, s_{23}, s_{34},
s_{45}\}$ in the rescaled coordinates of Section~6.3 of
Ref.~\cite{Bendle:2019csk} ($s_{12} \to 1$), with
$c_2 = s_{23}/s_{12}$, $c_3 = s_{34}/s_{12}$, $c_4 = s_{45}/s_{12}$, $c_5 = s_{15}/s_{12}$.  The same matrix
is used as the main reduction benchmark of
Ref.~\cite{Boehm:2020ijp}, which reports a $\sim$$25\times$ size
reduction of its partial-fraction form, and as the benchmark of
Section~5.3 of Ref.~\cite{MultivariateApart} on the
{\tt MultivariateApart} package.
The active variables we decompose against are
$\{\varepsilon, c_2, c_3, c_4, c_5\}$.  The parity-odd
Levi-Civita prefactor $\varepsilon_5$ (Eq.~4.7 of
Ref.~\cite{Boehm:2020ijp}) passes through as a spectator.  Since
Phase 1 is denominator-driven and does not touch the numerator,
we replace every cell's numerator by 1 before the benchmark run,
so that the recorded times reflect the partial-fraction work
itself rather than the size or shape of the numerator that
happens to sit in that cell of the original matrix.

\begin{figure}[!htbp]
    \centering
    \includegraphics[width=0.95\textwidth]{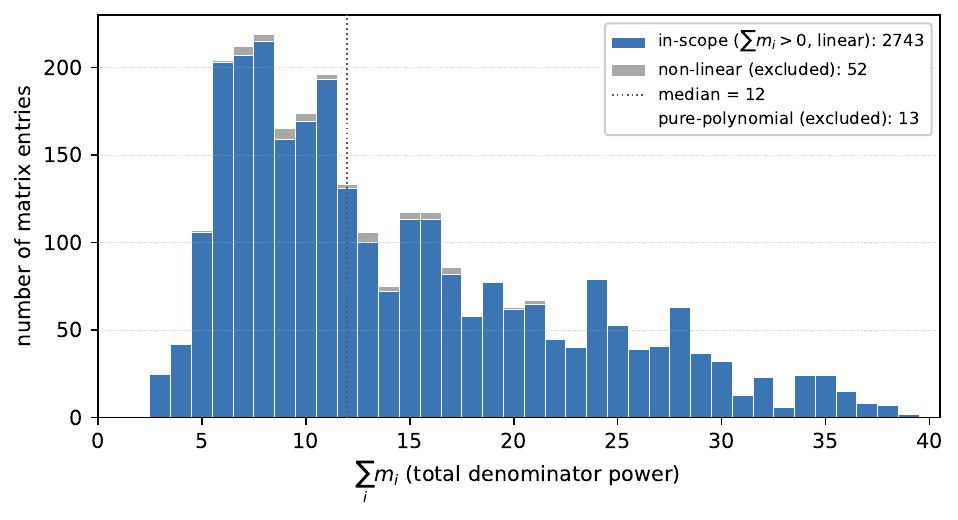}
    \caption{\label{fig:dpentagon_summ_histogram}
    Distribution of the total denominator power $\sum_i m_i$ across
    the $26 \times 108$ dpentagon dlog-basis IBP matrix of
    Ref.~\cite{Bendle:2019csk}: 2743 in-scope cells (blue), 52
    non-linear in the active variables (gray cap, excluded), and
    13 pure-polynomial entries (counted in the legend, also
    excluded).  Every in-scope cell of this matrix has
    $\sum_i m_i \le 39$.}
\end{figure}

\paragraph{Scope of the benchmark}
The 2808 cells of the matrix split into three disjoint groups. 
52 cells contain a denominator factor non-linear in the active variables (the Gram determinant
$G(1,2,3,4) = -16\,(\epsilon_{\mu\nu\rho\sigma}\,k_1^{\mu}k_2^{\nu}
k_3^{\rho}k_4^{\sigma})^2$ of the dlog basis,
concentrated in columns 49 and 63); these fall outside the algorithm of
Section~\ref{sect:algorithm} and are excluded from the timing.  A further
13 cells have no denominator that is linear in the active 
variables (literal zeros
or constants in those variables). Phase~1 has no
work to do and any measurement would only record \texttt{FORM} startup overhead; these are excluded as well.  The
2743 remaining cells are \emph{in scope} (at least one denominator linear in the active variables), and 
$2743 + 52 + 13 = 2808$.
Figure~\ref{fig:dpentagon_summ_histogram} shows the distribution of
the total denominator power $\sum_i m_i$ across the matrix, together
with the two excluded groups.

What stands out in
Figure~\ref{fig:dpentagon_summ_histogram} is that the in-scope
distribution runs from $\sum_i m_i = 3$ to $39$, so this matrix
covers a continuous range of input weights from very small to
moderately heavy.  We run the cache and the determinant branches on
every in-scope cell and compare them entry by entry below.

\begin{figure}[!htbp]
    \centering
    \includegraphics[width=0.95\textwidth]{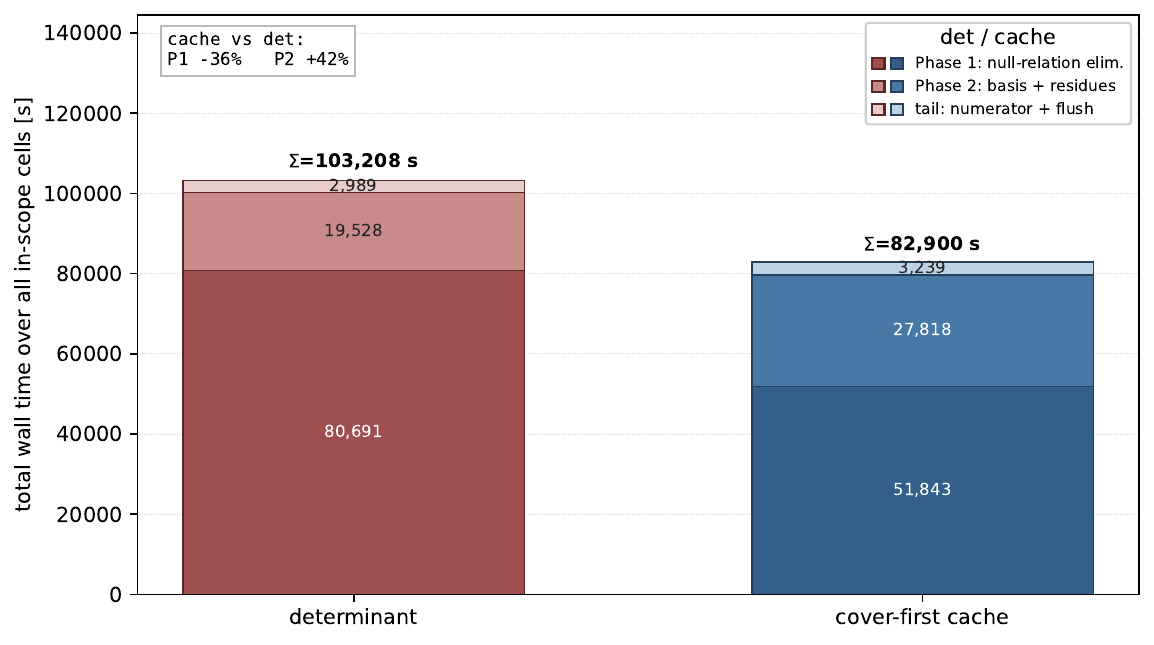}
    \caption{\label{fig:dpentagon_phase_totals}
    Total wall time over the 2743 in-scope dpentagon cells, split into
    Phase~1 (null-relation elimination), Phase~2 (basis identification
    and residues) and the numerator/flush tail, for the determinant
    and cover-first cache branches.  Times are the in-\texttt{FORM}
    \texttt{PHASEMARK} wall sums of the \texttt{-W} instrumented run;
    the three components sum to the per-mode total reported in
    Table~\ref{tab:dpentagon_totals}.}
\end{figure}

\begin{figure}[!htbp]
    \centering
    \includegraphics[width=0.95\textwidth]{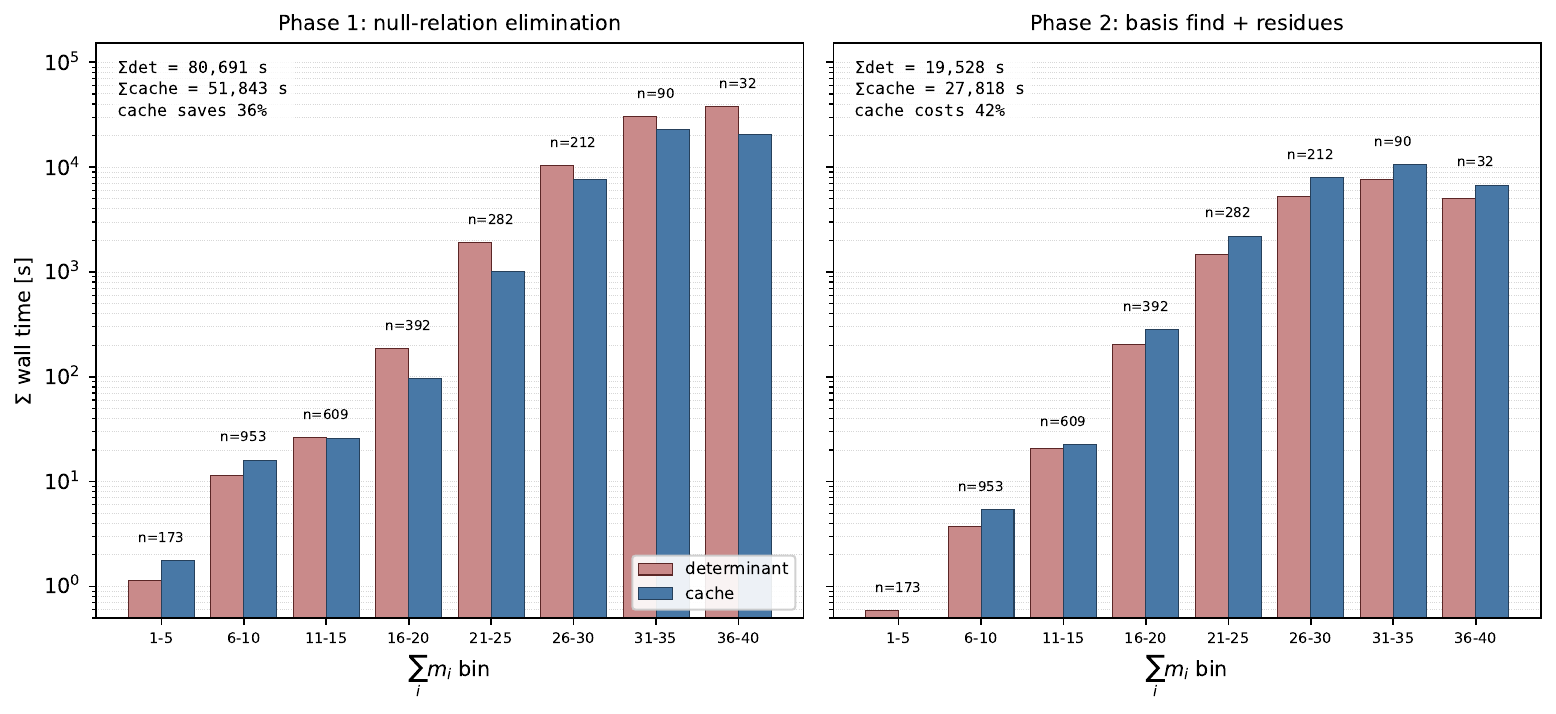}
    \caption{\label{fig:dpentagon_phase_perbin}
    Per-$\sum_i m_i$-bin wall time of the determinant branch (red) and
    the cover-first cache branch (blue), split into Phase~1 (left,
    null-relation elimination) and Phase~2 (right, basis
    identification and residues), over the 2743 in-scope dpentagon
    cells.  Bars are bin totals on a log scale; the cell count is
    annotated above each bin's bar pair.  The cache branch wins
    Phase~1 on the heavy tail and loses Phase~2 throughout, but the
    Phase~1 saving dominates the net.}
\end{figure}

\paragraph{Measurement protocol}
Each in-scope cell is written as a self-contained \texttt{FORM}
script with its numerator replaced by~1 as described above.  The
Phase~1 flag is fixed for each run, once with
\texttt{APuseGlobalNullRelations\,=\,0} (determinant branch) and
once with \texttt{APuseGlobalNullRelations\,=\,1} (cover-first
cache branch).  Wall-clock time is taken from
\texttt{/usr/bin/time -f "\%e"} around each \texttt{FORM}
invocation, capturing the whole process: \texttt{FORM} startup,
parse, the Phase~1 + Phase~2 pipeline, and shutdown. The per-phase numbers in 
Figures~\ref{fig:dpentagon_phase_totals}-\ref{fig:dpentagon_phase_perbin} come from a separate -W/PHASEMARK instrumented
run, whose whole-process total agrees with this /usr/bin/time measurement. The
benchmarks were run on an AMD Ryzen~9 5900XT workstation (16
physical cores, 128~GByte of DRAM), with the CPU clock overclocked
from its stock 3.3~GHz to 4.6~GHz and the DRAM access frequency
from 2666~MHz to 3200~MHz; \texttt{tform} was invoked with 16
workers throughout.

\begin{figure}[!htbp]
    \centering
    \includegraphics[width=0.85\textwidth]{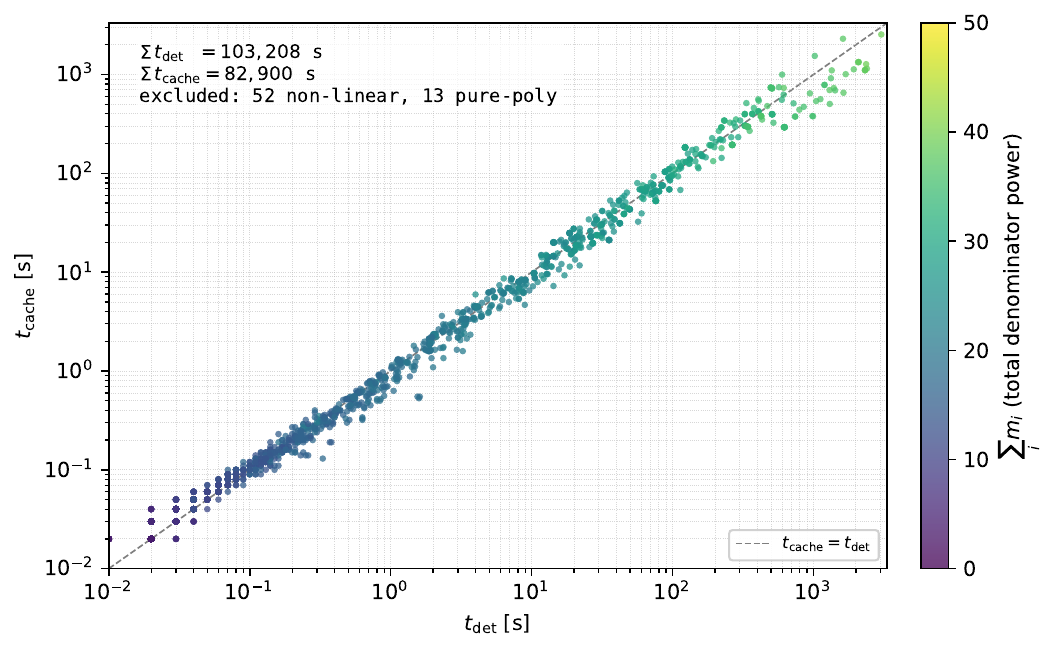}
    \caption{\label{fig:dpentagon_scatter}
    Per-cell wall time of the cover-first cache branch against
    the determinant branch on the 2743 in-scope dpentagon cells of
    Ref.~\cite{Bendle:2019csk}, colored by total denominator power
    $\sum_i m_i$.  The dashed diagonal $y=x$ marks where the two
    branches cost the same per cell; points below it are decomposed
    faster by the cache branch.  Per-mode
    totals (cf.~Table~\ref{tab:dpentagon_totals}): the cache branch
    saves $\sim$20\% of total wall time against the determinant
    branch.}
\end{figure}

\paragraph{Main finding}
Figure~\ref{fig:dpentagon_scatter} compares the two branches
entry by entry on the full in-scope population (all 2743 cells,
no subsampling), on a log-log scale with the cache wall time on
the $y$-axis and the determinant wall time on the $x$-axis.  Each
point is one matrix cell.  Color encodes $\sum_i m_i$.  The
dashed diagonal $y = x$ separates cache wins (below) from
determinant wins (above).

Three things stand out.  First, most of the points sit near the
noise floor: the per-cell median is around $60$~ms in both branches,
set by \texttt{FORM} startup, with Phase 1 doing very little on
small-$\sum_i m_i$ entries.  Second, the off-diagonal points all
go one way: the high-$\sum_i m_i$ cells (the green-teal
upper-right cluster) sit \emph{below} the diagonal, where the
cache branch wins.  No entries sit visibly above the diagonal in
the heavy tail.  Third, the per-mode total wall times are dominated by
the heavy tail, so the per-cell median tells us very little about
the total.

The aggregate numbers are summarized in
Table~\ref{tab:dpentagon_totals}.  The cache branch saves
$\sim$20\% of total wall time against the determinant branch on the
full matrix, with the heaviest in-scope cells (where the bulk of
either branch's wall time lives) sitting where the cache branch
pulls ahead most clearly.

\begin{table}[H]
\centering
\renewcommand{\arraystretch}{1.3}
\begin{tabular}{|l|r|r|r|}
\hline
\textbf{Mode} & \textbf{$\Sigma$ wall [s]} & \textbf{max [s]} & \textbf{median [s]} \\
\hline\hline
determinant & 103\,208 & 3\,027 & 0.06 \\
\hline
cache       &  82\,900 & 2\,544 & 0.06 \\
\hline
\end{tabular}
\caption{\label{tab:dpentagon_totals}
Per-mode wall-time aggregates over the full 2743-cell in-scope
population of the dpentagon matrix of
Ref.~\cite{Bendle:2019csk}.  The cache branch's maximum per-cell
wall time ($2\,544$~s) is about $\sim$16\% below the determinant
branch's ($3\,027$~s), and the bulk of the $\sim$20\% total saving
comes from the heavy tail.}
\end{table}

\begin{figure}[!htbp]
    \centering
    \includegraphics[width=0.95\textwidth]{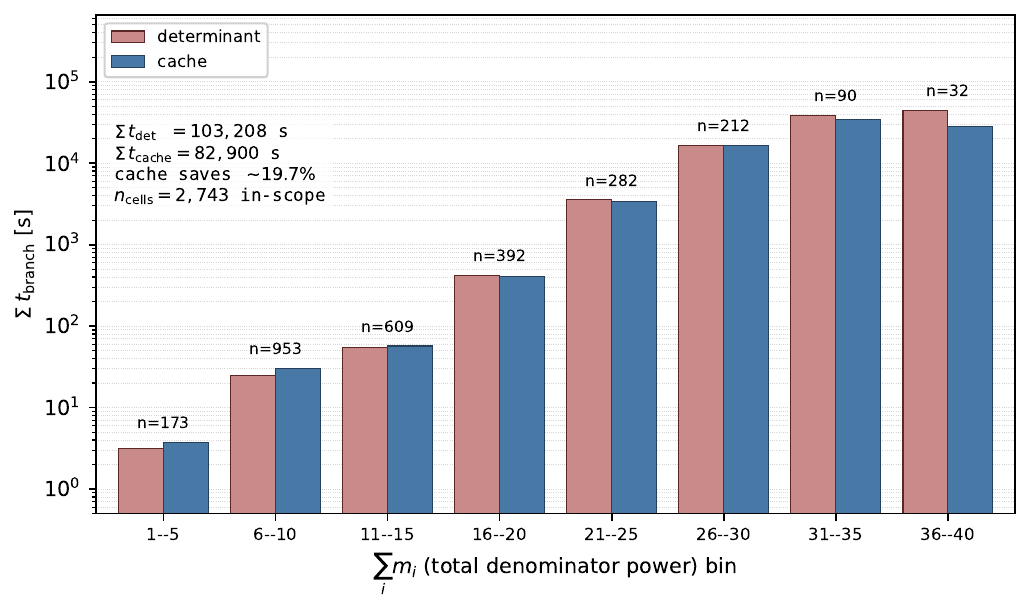}
    \caption{\label{fig:dpentagon_per_bin}
    Per-$\sum_i m_i$-bin wall time of the determinant branch (red)
    and the cover-first cache branch (blue), aggregated over the
    2743 in-scope dpentagon cells of
    Ref.~\cite{Bendle:2019csk}.  Bars are bin totals on a log
    scale.  The cell count is annotated above each bin's bar pair.}
\end{figure}

\paragraph{Per-bin breakdown}
Figure~\ref{fig:dpentagon_per_bin} resolves where the cache
saving comes from.  Each width-5 bin of $\sum_i m_i$ contributes
two side-by-side bars, the determinant total (red) and the cache
total (blue), on a log $y$-axis.  Three observations:
(i) the bins with $\sum_i m_i \le 20$ together contribute only
$\sim 500$~s ($\sim 0.5\%$) of either branch's total, and they
are the noise-floor cells of the scatter plot.
(ii) The bins with $\sum_i m_i$ between 21 and 30 are nearly flat
between the determinant and cache branches (the cache branch even
loses about 25~s in the bin with $\sum_i m_i$ between 26 and 30), so the net saving in
that range is only of order $100$~s.
(iii) The heavy tail with $\sum_i m_i$ between 31 and 40 contains
only 122 of the 2743 cells but accounts for $\sim 80\%$ of the
determinant total wall time and $\sim 76\%$ of the cache total wall time,
and this is where the cache branch pulls ahead.  The 32 cells with $\sum_i m_i$ between
36 and 40 alone save $\sim 15\,900$~s, with a further
$\sim 4\,300$~s saved by the 90 cells with $\sum_i m_i$ between 31
and 35, together accounting for essentially the whole
$\sim 20\,000$~s saving.  On this benchmark the cache branch is therefore 
the more economic choice
in aggregate, with the saving concentrated in the heavy tail; 
on the lighter bins
the two branches are within a few per cent and the determinant branch is occasionally
faster (e.g. the 26 to 30 bin).

\paragraph{Phase 1 vs Phase 2 breakdown}
The wall times reported so far combine the two phases of the
algorithm.  To see where each branch spends its time, we re-ran the
full in-scope population with per-phase instrumentation:
\texttt{tform} with the \texttt{-W} flag (wall-clock module timings)
and \texttt{\#message PHASEMARK} boundaries emitted after the parse,
after Phase~1 (null-relation elimination) and after Phase~2 (basis
identification and residues), so that the wall time of each phase is
recorded separately.

Figure~\ref{fig:dpentagon_phase_totals} splits the totals by phase.
Two facts stand out.  First, Phase~1 dominates: it accounts for
$\sim$81\% of the combined Phase~1\,+\,Phase~2 wall time in the
determinant branch and $\sim$65\% in the cache branch.  Second, the
cover-first cache's entire advantage lives in Phase~1, where it cuts
the total from $\sim$80\,700~s to $\sim$51\,800~s ($-36\%$); in
Phase~2 it is in fact $\sim$42\% \emph{slower} ($\sim$19\,500~s
against $\sim$27\,800~s).  Because Phase~1 so dominates, the net effect
is still the $\sim$20\% overall saving of
Table~\ref{tab:dpentagon_totals}.

Figure~\ref{fig:dpentagon_phase_perbin} resolves this per
$\sum_i m_i$ bin.  In Phase~1 (left panel) the cache branch is at or
below the determinant branch on every populated bin and pulls clearly
ahead on the heavy tail, mirroring the total per-bin breakdown of
Figure~\ref{fig:dpentagon_per_bin}.  In Phase~2 (right panel) the
ordering reverses: the cache branch is consistently the more expensive
of the two.  The cover-first cache therefore trades a small,
broadly-distributed Phase~2 overhead for a large Phase~1 saving on the
heavy cells that dominate the total wall time.

\FloatBarrier

\section{Conclusions and outlook}
\label{sect:conclusions}

We have presented {\tt LinApart3}, a multivariate partial fraction decomposition algorithm for rational functions with linear denominators. The algorithm exploits the geometry of the hyperplane arrangement defined by the denominators, replaces the polynomial-ideal computations of the Gröbner-basis method and circumvents the iterative nature of Leinartas' algorithm with linear algebra and
multivariate residue extraction.

The algorithm proceeds in four steps. First, null-relations among the denominators are used to construct one-operators that recursively eliminate linearly dependent
denominators. Second, we determine all basis denominators. Third, variable-dependent numerators are expanded in denominator coordinates to produce constant numerators. Fourth, partial fraction coefficients are extracted via the Grothendieck residue formula in denominator space, the direct multivariate generalization of the univariate Laurent series residues used in {\tt LinApart}~\cite{LinApart} and {\tt LinApart2}~\cite{LinApart2}.

The resulting decomposition guarantees at most $n$ distinct denominators per term for $n$ variables, uses only denominators from the original expression, so no spurious singularities appear. The algorithm is also independent of variable ordering, and when the input denominators carry no null relations the output is unique. Because the basis residues are independent, they can be computed in parallel, leaving room for further runtime optimization.

We have compared the performance and sensitivity of {\tt LinApart3} against three alternative methods: the Leinartas and Gr\"obner-basis methods implemented in our package, and {\tt MultivariateApart}. The main findings are:
\begin{itemize}
    \item {\tt LinApart3} exhibits the most favorable scaling with the number of denominator factors: over the tested range its growth is roughly polynomial while the other methods grow much faster (consistent with exponential behavior). This is the strongest argument for using it, since the difference can amount to orders-of-magnitude savings even at the modest (about 20) denominator counts tested.
    \item The algorithm is insensitive to the number of spectator variables: parameters that do not participate in the decomposition do not affect the runtime. This is in sharp contrast to the Gröbner-basis method, where each additional parameter in the coefficient ring sharply increases the cost of the polynomial-ideal computation.
    \item Our algorithm also shows only a mild dependence on the number of partial-fraction variables and was the best performing of the tested methods for fractions with many high-multiplicity denominators. These are particularly important features for high-order perturbative Quantum Chromodynamics calculations.
    \item Based on our benchmarks the {\tt LinApart3} algorithm can also handle fractions with numerators and is competitive with the Gröbner-method there.
    \item It does, however, scale poorly with the number of null relations, which is its main bottleneck. 
\end{itemize}

Based on these benchmarks we believe that {\tt LinApart3} is particularly well-suited to tackle expressions arising in perturbative quantum field theory, where rational functions of many kinematic invariants and the dimensional regulator must be decomposed with respect to a chosen subset of variables, and the introduction of spurious singularities is unacceptable.

Several directions for future development are worth pursuing.

\paragraph{Circumventing null-relation elimination}
As we have seen, {\tt LinApart3} shows the worst performance in the case when linearly dependent denominators are present. It could be that someone introduces such constraints during the basis identification that could circumvent this bottleneck and directly give the appropriate bases, thus eliminating this major slow-down effect. 

\paragraph{Optimization of the null-relation elimination order}
The current implementation uses a greedy heuristic (highest multiplicity first) to select which denominator to eliminate at each step. This does not guarantee that the total number of terms in the output is minimized. Exploring alternative elimination strategies could reduce the output size and improve performance for expressions with many null-relations. In case of the \texttt{FORM} implementation we gave alternate strategies to the user to experiment with which approach better suits the problem at hand.

\paragraph{Extension to non-linear denominators}
{\tt LinApart3} is restricted to denominators that are linear in the decomposition variables. Extending the residue-based approach to handle irreducible quadratic or higher-degree denominators, in analogy with the univariate extension from {\tt LinApart} to {\tt LinApart2}, would broaden the applicability of the method. The key challenge is the coordinate transformation to denominator space: for non-linear denominators, the intersection locus is no longer a single point but a higher-dimensional variety, and the residue computation becomes correspondingly more involved and the invertibility of the transformation into denominator space is questionable.

\paragraph{Integration with finite-field methods}
In state-of-the-art multi-loop calculations, rational functions are often reconstructed from finite-field samples. Combining {\tt LinApart3} with finite-field sampling and rational reconstruction techniques could enable partial fraction decomposition of expressions that are too large to handle symbolically, by decomposing the sampled values and reconstructing the partial-fractioned coefficients directly.

\paragraph{Connection to intersection theory}
The residue extraction central to {\tt LinApart3} localizes at vertices of the hyperplane arrangement, points where $n$ hyperplanes $\{B_i(\mathbf{x})=0\}$ meet, which is precisely the localization that appears in the global residue theorem underlying intersection-number computations for Feynman integrals~\cite{Mastrolia:2018uzb, Frellesvig:2019uqt}. In both cases, the coordinate transformation to denominator space maps each vertex to the origin, and the partial fraction coefficients (respectively, the local intersection numbers) are extracted by residue calculus at that point. The essential difference is that intersection theory requires a twisted cohomology structure: the integrand carries a multivalued factor $u(\mathbf{x})=\prod_i D_i(\mathbf{x})^{\alpha_i}$ with exponents $\alpha_i$ depending on the dimensional regulator, and the relevant residues are computed with respect to the covariant derivative $\nabla = d + d\!\log u\,\wedge$ rather than the ordinary de~Rham differential. Whether the efficient hyperplane coordinate transformations and basis enumeration of {\tt LinApart3} can be adapted to the twisted setting (for instance, by treating the twist exponents perturbatively in $\epsilon$ or by fitting them with some finite-field method) is an open question whose resolution could offer a new route to faster intersection-number evaluations.

\section*{Acknowledgements}

We are grateful to Sven Moch and Oliver Schnetz for several useful discussions. We acknowledge support from the ERC
Advanced Grant 101095857 Conformal-EIC. A.K. is supported by the
University of Debrecen Program for Scientific Publication.


\bibliographystyle{apsrev4-1}
\bibliography{refs}

\end{document}